\documentclass[preprint,12pt]{elsarticle}

\usepackage{amssymb}
\usepackage{amsmath}
\usepackage{setspace}
\usepackage{amsthm}
\usepackage{color}

\usepackage{lscape}

\usepackage{graphics}
\usepackage{graphicx}

\usepackage{caption}
\usepackage{subcaption}

\usepackage[boxed, algoruled, linesnumbered, inoutnumbered, noend]{algorithm2e}
\usepackage{mathrsfs}
\usepackage{multicol}
\usepackage{multirow}
\usepackage{empheq}
\usepackage{ctable}

\biboptions{numbers,sort}

\journal{Computers \& Operations Research}
\onehalfspace

\begin{document}

\begin{frontmatter}



\title{A Variable Fixing Heuristic with Local Branching for the Fixed Charge Uncapacitated Network Design Problem with User-optimal Flow}


\author[UFF,UAPV]{Pedro Henrique Gonz\'{a}lez}\ead{pegonzalez@ic.uff.br}
\author[UFF]{Luidi Simonetti}\ead{luidi@ic.uff.br}
\author[UAPV]{Philippe Michelon}\ead{philippe.michelon@univ-avignon.fr}
\author[UFF]{Carlos Martinhon}\ead{mart@dcc.ic.uff.br}
\author[UFF]{Edcarllos Santos}\ead{esantos@ic.uff.br}

\address[UFF]{Instituto de Computa\c{c}\~{a}o, Universidade Federal Fluminense, Niter\'{o}i, Brazil}
\address[UAPV]{Laboratoire d'Informatique d'Avignon, Universit\'{e} d'Avignon et des Pays de Vaucluse, Avignon, France}

\begin{abstract}
This paper presents an iterated local search for the fixed-charge uncapacitated network design problem with user-optimal flow (FCNDP-UOF), which concerns routing multiple commodities from its origin to its destination by designing a network through selecting arcs, with an objective of minimizing the sum of the fixed costs of the selected arcs plus the sum of variable costs associated to the flows on each arc. Besides that, since the FCNDP-UOF is a bilevel problem, each commodity has to be transported through a shortest path, concerning the edges length, in the built network. The proposed algorithm generate a initial solution using a variable fixing heuristic. Then a local branching strategy is applied to improve the quality of the solution. At last, an efficient perturbation strategy is presented to perform cycle-based moves to explore different parts of the solution space. Computational experiments shows that the proposed solution method consistently produces high-quality solutions in reasonable computational times.
  
\end{abstract}


\end{frontmatter}



\section{INTRODUCTION}
\label{sec:introduction}
Due to the continuous development of society, increasing quantities of commodities have to be transported in large urban centers. Therefore, network design problems arise as tools to support decision-making, aiming to meet the need of finding efficient ways to perform the transportation of each commodity from its origin to its destination.
In the Fixed Charge Network Design Problem (FCNDP), a subset of edges is selected from a graph, in such a way that a given set of commodities can be transported from their origins to their destinations. The main objective is to minimize the sum of the fixed costs (due to selected edges) and variable costs (depending on the flow of goods on the edges). In addition, fixed and variable costs can be represented by linear functions and arcs are not capacitated. Belonging to a large class of network design problems, the FCNDP has several variations such as shortest path problem, minimum spanning tree problem, vehicle routing problem, traveling salesman problem and Steiner problem in graph \cite{Magnanti1984}. For generic network design problem, such as FCNDP, numerous applications can be found \cite{Boesch1976,Boyce1980,Mandl1981}, thus, mathematical formulations for the problem may also represent several other problems, like problems of communication, transportation, sewage systems and resource planning. It also appears in other contexts, such as flexible production systems \cite{Kimemia1978} and automated manufacturing systems \cite{Graves1983}. Finally, network design problems arise in many vehicle fleet applications that do not involve the construction of physical facilities, but rather model decision problems such as sending a vehicle through a road or not \cite{Magnanti1981,Simpson1969}.\\
This work addresses a specific variation of FCNDP, called Fixed-Charge Uncapacitated Network Design Problem with User-optimal Flows (FCNDP-UOF), which consists of adding multiple shortest path problems to the original problem. The  FCNDP-UOF involves two distinct agents acting simultaneously rather than sequentially when making decisions. On the upper level, the leader ($1^{st}$ agent) is in charge of choosing a subset of edges to be opened in order to minimize the sum of fixed and variable costs. In response, on the lower level, the follower ($2^{nd}$ agent) must choose a set of shortest paths in the network, through which each commodity will be sent. The effect of an agent on the other is indirect: the decision of the follower is affected by the network designed on the upper level, while the leader's decision is affected by variable costs imposed by the routes settled in the lower level. The inclusion of shortest path problem constraints in a mixed integer linear programming is not straightforward. Difficulties arise both in modeling and designing efficient methods.\\
The FCNDP-UOF problem appears in the design of a network for hazardous materials transportation \cite{Amaldi2011:HTN:2040817.2040858,Erkut2008,Erkut2007,Kara2004}. Particularly for this kind of problem, the government defines a selection of road segments to be opened/closed to the transportation of hazardous materials assuming that the shipments in the resulting network will be done along shortest paths. In hazardous materials transportation problems, roads selected to compose the network have no costs, but the goverment wants to minimize the population exposure in case of an incident during a dangerous-goods transportation. This is a particular case of the FCNDP-UOF problem where, from a mathematical point of view, the fixed costs are equal to zero.\\
Several variants of the FCNDP-UOF can be seen on \cite{Amaldi2011:HTN:2040817.2040858,Billheimer1973,Erkut2008,Erkut2007,Gonzalez2013,Kara2004,Mauttone2008}  and have been treated as part of larger problems in some applications on \cite{Holmberg2004}. The work presented by Bilheimer and Grey \cite{Billheimer1973} formally defines the FCNDP-UOF.  Both Erkut et al. \cite{Erkut2007} and Kara et al. \cite{Kara2004} work focus on exact methods, presenting a mathematical formulation and several metrics for the hazardous materials transportation problem. At Mauttone et al. \cite{Mauttone2008}, not only was presented a different model, but also a Tabu Search for the FCNDP-UOF. Both, Amaldi et al. \cite{Amaldi2011:HTN:2040817.2040858} and Erkut et al. \cite{Erkut2008} presented heuristic approaches to deal with the hazardous materials transportation problem. At last, Gonzalez et al. \cite{Gonzalez2013}, presented an extension of the model proposed by Kara and Verter \cite{Kara2004} and also a GRASP.\\ 
According to \cite{Johnson1978,Wong1978}, the simplest versions of network design problems are $\mathcal{NP}$-hard and even the task of finding feasible solutions (for problems with budget constraint on the fixed cost) is extremely complex \cite{Wong1980}. Therefore, heuristics methods are presented as a good alternative in the search for good solutions. Knowing that, this work proposes an Iterated Local Search \cite{Lourenco2010} for the FCNDP-UOF\\
This text is organized as follows. In Section 2, we start by describing the problem followed by a bi-level and an one-level formulation, presented on \cite{Mauttone2008}. Then in Section 3 we present our solution approach. Section 4 reports on our computational experiments. At last, in Section 5 the conclusion and future works are presented.

\section{GENERAL DESCRIPTION OF FCNDP-UOF}

In this section we describe the problem and present a bi-level and an one-level formulation for the FCNDP-UOF proposed respectively by \cite{Colson2005,Mauttone2008} for the FCNDP-UOF.\\
The basic structures to create a network are a set of nodes $V$ that represents the facilities and a set of uncapacitated and undirected edges $E$ representing the connection between installations. Furthermore, the set $K$ is the set of commodities to be transported over the network, and these commodities may represent physical goods as raw material for industry, hazardous material or even people. Each commodity $ k \in K $, has a flow to be delivered through a shortest path between its source $o(k)$ and its destination $d(k)$.  The formulation presented here works with variants presenting commodities with multiple origins and destinations, and for treating such a case, it is sufficient to consider that for each pair ($o(k),d(k)$), there is a new commodity resulting from the dissociation of one into several commodities. \\

\subsection{Mathematical Formulation}

This subsection presents a few definitions in order to make easier the understanding of the problem.\\
The model for FCNDP-UOF has two types of variables, one for the construction of the network and another representing the flow. Let $ y_ {ij} $ be a binary variable, we have that $ y_{ij} =  1$ if the edge $ [i,j] $ is chosen as part of the network and $ y_{ij} =  0$ otherwise. In this case, $ x^{k}_{ij} $ denotes the commodity $k$'s flow through the arc $(i,j)$. Although the edges have no direction, they may be referred to as arcs, because each commodity flow is directed. Treating $ y = (y_ {ij}) $ and $ x^k =(x^{k}_{ij})$, respectively, as vectors of active edge and flow variables, mixed integer programming formulations can be elaborated.

\subsection*{List of Symbols}
\begin{tabular}{lll}
$V$ & Set of nodes.  \\
$E$ & Set of admissible edges.\\
$K$ & Set of commodities.\\
$A^{E}$ & Set of arcs obtained by bidirecting the edges
in E. \\
$\mathcal{G}$ & Associated graph $G(V,E)$.\\
$\delta^{+}_{i}$ & Set of all arcs leaving node $i$. \\
$\delta^{-}_{i}$ & Set of all arcs arriving at node $i$.  \\
$c_a$ & Length of the arc $a$. \\
$e(a)$ & Edge $e$ related to the arc $a$. \\
$o(k)$ &  Origin node for commodity $k$. \\
$d(k)$ &  Destiny node for commodity $k$. \\
$g^{k}_{ij}$ &  Variable cost of transporting commodity \\
	& $k$ through the arc $(i,j) \in A^E $. \\
$f_{ij}$ &  Fixed cost of opening the edge $[i,j] \in E$.\\
$y_{ij}$ &  Indicates whether edge $[i,j]$ belongs in the solution. \\
$x^{k}_{ij}$ & Indicates whether commodity $k$ passes through \\
	& the arc $(i,j)$. \\
\end{tabular}

\subsection{Bi-level Formulation}
In FCNDP-UOF, differently from the basic FCNDP, each commodity $k \in K$ has to be transported through a shortest path between its origin $o(k)$ and its destination $d(k)$,  forcing the addition of new constraints to the general problem. Besides selecting a subset of $ E $ whose sum of fixed and variable costs is minimal (leading problem), in this variation, we also have to garantee the shortest path constraints for each commodity $k \in K$ (follower problem). The FCNDP-UOF belongs to the class of $\mathcal{NP}$-hard problems and can be modeled as a bi-level mixed integer programming problem \cite{Colson2005}, as follows:

\footnotesize
\begin{empheq}{align}
	\min \quad & \displaystyle\sum\limits_{e \in E} f_{e}y_{e} + \sum\limits_{k \in K}\sum\limits_{(i,j) \in A^{E}} g_{ij}^k x_{ij}^k  \nonumber\\
	\textrm{s.t.} \quad &  y_{e} \in \{ 0,1 \}, & \forall e \in E, \label{y} 
\end{empheq}

\normalsize
\noindent where $x_{ij}^k$ is a solution of the problem:

\footnotesize
\begin{empheq}{align}
	\min \quad & \displaystyle \sum\limits_{k \in K}\sum\limits_{{a=(i,j)} \in A^{E}}c_{a}x_{ij}^k  \nonumber\\ 
	\textrm{s.t.}\quad & \displaystyle\sum\limits_{(i,j) \in \delta ^+(i)} x_{ij}^k - \sum\limits_{(i,j) \in \delta ^-(i)} x_{ji}^k  = b_i^k, & \forall i \in V, \forall k \in K, \label{Fluxo2} \\
	&  x_{ij}^k + x_{ji}^k \leq y_{e}, &\forall e=[i,j] \in E, \forall k \in K, \label{xmy2} \\
	&  x_{ij}^k \geq 0, &\forall (i,j) \in A^E, \forall k \in K. \label{x2}
\end{empheq}
\normalsize
\noindent where:
\footnotesize
 \begin{empheq}{align}
 b_i^k  = \displaystyle\qquad\left\{\begin{array}{rl} -1 & \textrm{if } i = d(k),\\ 1 & \textrm{if } i = o(k),\\ 0 & \textrm{otherwise.}\\
\end{array}\right. \nonumber
\end{empheq}
\normalsize
According to constraints (\ref{y})-(\ref{x2}), we can notice that the set of constraints (\ref{y}) ensures that the vector of variables $y$ assume only binary values. In (\ref{Fluxo2}), we have  flow conservation constraints. Constraints (\ref{xmy2}) do not allow flow into arcs whose corresponding edges are closed. Finally, (\ref{x2}) imposes the non-negativity restriction of the vector of variables $x^k$. An interesting remark is that solving the follower problem is equivalent to solving $|K| $ shortest path problems independently.\\

\subsection{One-level Formulation}

The FCNDP-UOF can be formulated as a one-level integer programming problem replacing the objective function and the constraints defined by (\ref{Fluxo2})-(\ref{x2}) of the follower problem for its optimality conditions \cite{Mauttone2008}. This can be done by applying the fundamental theorem of duality and the complementary slackness theorem \cite{Bazaraa:2004:LPN:1062374}, as follows:\\
\footnotesize
\begin{empheq}{align}
	\min \quad & \displaystyle\sum\limits_{e \in E} f_{e}y_{e} + \sum\limits_{k \in K}\sum\limits_{(i,j) \in A^{E}} g_{ij}^k x_{ij}^k  \nonumber\\
	\textrm{s.t.}\quad & \displaystyle\sum\limits_{{(i,j)} \in \delta ^+(i)} x_{ij}^k - \sum\limits_{(i,j) \in \delta ^-(i)} x_{ji}^k  = b_i^k, & \forall i \in V, \forall k \in K, \\
	&  x_{ij}^k + x_{ji}^k \leq y_{e}, & \forall e=[i,j] \in E, \forall k \in K ,\\
	& \pi_i^k - \pi_j^k - \lambda_{e(a)}^{k} \leq c_{a} &\forall  a= ( i,j) \in A^{E}, k\in K, \label{dual1}\\
	& (y_e - x_{ij}^k - x_{ji}^k) \lambda_{e}^{k} = 0, & \forall e=[i,j] \in E, \forall k \in K, \label{dual2}\\
	& (c_a - \pi_i^k + \pi_j^k + \lambda_{e(a)}^{k})x_{ij}^{k} = 0, &  \forall  a= ( i,j) \in A^{E}, k\in K, \label{dual3}\\\
	& \lambda _{e}^{k} \geq 0, &\forall  e = [ i,j ] \in E, k\in K, \\
	& \pi_i^k \in \mathbb{R}, &\forall i \in V, \forall k \in K, \\
	&  x_{ij}^k \geq 0, &\forall (i,j) \in A^{E}, \forall k \in K, \\
	&  y_{e} \in \{ 0,1 \}, & \forall e \in E.
\end{empheq}
\normalsize
\noindent where:
\footnotesize
\begin{empheq}{align}
b_i^k  = \displaystyle\qquad\left\{\begin{array}{rl} -1 & \textrm{if } i = d(k),\\ 1 & \textrm{if } i = o(k),\\ 0 & \textrm{otherwise. }\\
\end{array}\right. \nonumber
\end{empheq}
\normalsize
A disadvantage of this new formulation is the loss of linearity of the model. To bypass this problem, a Big-M linearization may be used. After it, one can write the model as a one-level mixed integer linear programming problem, as follows:
\footnotesize
\begin{empheq}{align}
	\min \quad & \displaystyle\sum\limits_{e \in E} f_{e}y_{e} + \sum\limits_{k \in K}\sum\limits_{(i,j) \in A^{E}} g_{ij}^k x_{ij}^k  \nonumber\\
	\textrm{s.t.}\quad & \displaystyle\sum\limits_{{(i,j)} \in \delta ^+(i)} x_{ij}^k - \sum\limits_{(i,j) \in \delta ^-(i)} x_{ji}^k  = b_i^k, & \forall i \in V, \forall k \in K, \label{0}\\
	&  x_{ij}^k + x_{ji}^k \leq y_{e}, & \forall e=[i,j] \in E, \forall k \in K \label{1} \\
	& \pi_i^k - \pi_j^k - \lambda_{e(a)}^{k} \leq c_{a} &\forall  a= ( i,j) \in A^{E}, k\in K, \label{dual6}\\
	& \lambda_{e}^{k} + M_{e}y_{e} - M_{e}x_{ij}^{k} - M_{e}x_{ji}^{k} \leq M_{e}, & \forall e=[i,j] \in E, \forall k \in K, \label{dual4}\\
	& M_{e(a)}x_{ij}^{k} - \pi_i^k + \pi_j^k + \lambda_{e(a)}^{k} \leq M_{e(a)} - c_{a}, &  \forall  a= ( i,j) \in A^{E}, k\in K, \label{dual5}\\
	& \lambda _{e}^{k} \geq 0, &\forall  e = [i,j] \in E, k\in K, \label{2}\\
	& \pi_i^k \in \mathbb{R}, &\forall i \in V, \forall k \in K, \label{3}\\
	&   x_{ij}^k \in \{ 0,1 \}, &\forall (i,j) \in A^{E}, \forall k \in K, \\
	&  y_{e} \in \{ 0,1 \}, & \forall e\in E.
\end{empheq}
\normalsize
 \noindent where:
\footnotesize
\begin{empheq}{align}
b_i^k  = \displaystyle\qquad\left\{\begin{array}{rl} -1 & \textrm{if } i = d(k),\\ 1 & \textrm{if } i = o(k),\\ 0 & \textrm{otherwise. }\\
\end{array}\right. \nonumber
\end{empheq}
\normalsize
However, optimality conditions for the problem in the lower level are, in fact, the optimality conditions of the shortest path problem and they could be expressed in a more compact and efficient way if we consider Bellman's optimality conditions for the shortest path problem \cite{Ahuja:1993:NFT:137406} and using a simple lifting process \cite{LuigiDeGiovanni2004}.\\

\footnotesize
\begin{empheq}{align}
	\min \quad & \displaystyle\sum\limits_{e \in E} f_{e}y_{e} + \sum\limits_{k \in K}\sum\limits_{(i,j) \in A^{E}} g_{ij}^k x_{ij}^k  \nonumber \\
	\textrm{s.t.}\quad & \displaystyle\sum\limits_{{(i,j)} \in \delta ^+(i)} x_{ij}^k - \sum\limits_{(i,j) \in \delta ^-(i)} x_{ji}^k  = b_i^k, & \forall i \in V, \forall k \in K, \label{EQF}\\
	&  x_{ij}^k + x_{ji}^k \leq y_{ij}, & \forall e=[i,j] \in E, \forall k \in K ,\label{aqui}\\
	& \pi_i^k - \pi_j^k \leq M_{e(a)} - y_{e(a)}(M_{e(a)} - c_a) - 2c_ax_{ji}^k, &\forall  a= (i,j) \in A^{E}, k\in K, \label{dual}\\
	& \pi_{d(k)} ^{k} =0 , & \forall k \in K, \label{rosa2}\\
	& \pi_i^k \geq 0, &\forall i \in \setminus \{ d(k) \}, \forall k \in K, \label{rosa1} \\
	&  x_{ij}^k \in \{ 0,1 \}, &\forall (i,j) \in A^E, \forall k \in K, \label{xdom}\\
	&  y_{e} \in \{ 0,1 \}, & \forall e\in E.\label{FM}
\end{empheq}

\normalsize
 \noindent where:
\footnotesize
\begin{empheq}{align}
b_i^k  = \displaystyle\qquad\left\{\begin{array}{rl} -1 & \textrm{if } i = d(k),\\ 1 & \textrm{if } i = o(k),\\ 0 & \textrm{otherwise. }\\
\end{array}\right. \nonumber
\end{empheq}
\normalsize
The variables $ \pi_{i}^k $, $ k \in K $, $ i \in V $, represent the shortest distance between vertex $i$ and vertex $ d(k) $. Then we define that $ \pi_{d(k)}^k$ will always be equal zero. Assuming that constraints (\ref{aqui}), (\ref{xdom}) and (\ref{FM}) are satisfied, it is easy to see that constraints (\ref{dual}) are equivalent to Bellman's optimality conditions for $|K|$ pairs $(o(k), d(k))$.

\section{SOLUTION APPROACH}

This section focuses on presenting the different methods developed in this work. First the Partial Decoupling Heuristic is introduced. Secondly a procedure to find a lower bound. After that a variable fixing heuristic that uses the previously explained methods. At last a Local Branching (used as Local Search) and a Ejection Cycle (used as Pertubation) are shown so a Iterated Local Search metaheuristics could be done.

\subsection{Partial Decoupling Heuristic}
The main idea of total decoupling heuristic for the FCNDP-UOF is dissociating the problem of building a network from the shortest path problem. This disintegration, as discussed in \cite{Erkut2008}, can provide worst results than when addressing both problems simultaneously. To work around this situation, the method uses what we call partial decoupling, where certain aspects of the follower problem are considered when trying to build a solution to the leading problem.\\
The Partial Decoupling Heuristic iterativily builds a network and then routes each commodity so a feasible solution can be built. In order to build the network the cost $\bar{f}_{e}^{k} ,\mbox{ } e \in E,\mbox{ } k \in K $ is defined:

\begin{equation}
	\bar{f}_{e}^{k} = \left\{\begin{array}{ll}
    f_{e} +\alpha \times g^{k}_{ij}+(1 - \alpha) \times c_e & \quad \text{if $y_{e}=0$,}\\
    \alpha \times g^{k}_{ij}+(1 - \alpha) \times c_e & \quad \text{otherwise.}\\
  \end{array} \right.
\end{equation}

\noindent Doing that we consider whether the edge is open or not, plus a linear combination of the variable cost and the length of the edge as the fixed cost. The $\alpha$ works as a scaling parameter of the importance of the $g^{k}_{ij}$ and $c_e$ values. In the beginning of the heuristic $\alpha$ prioritizes the variable cost ($g^{k}_{ij}$), while in the end it prioritizes the edge length ($c_e$). It is important to pay attention that $ g_{ij}^k = q^{k} \beta _{ij} $, where $ q^{k} $ represents the amount of commodity $ k $ to be transported and $ \beta _{ij} $ represents the shipping cost through the edge $ e = (i,j) $.\\
After building the network, another shortest path algorithm, using the edges length ($c_e$) as cost, is applied to take every commodity from its origin $o(k)$ to its destination $d(k)$ in the built network.\\
In order to put the scaling parameter $\alpha$ in good use, the method repeats $MaxIterDP$ times and at each iteration using a different value for $\alpha$.
The proposed algorithm is a small variation of the original Partial Decoupling Heuristic \cite{Gonzalez2013}. The procedure is further explained on Algorithm \ref{alg1}.
\pagebreak
\begin{algorithm}[htbp]

\SetKwInOut{Input}{Input}

\Input{$\gamma$, $K$, $\mathcal{G}$}
\KwData{$MinCost \leftarrow \infty$, $\alpha \leftarrow 1$, $y \leftarrow 0$, $ x \leftarrow 0$;}
\Begin{
  $\bar{K} \leftarrow K$;\\
  \For{$numIterDP \mbox{ in } 1 \ldots MaxIterDP$}{
  	\While{$ \bar{K} \neq \emptyset $}{
		$\hat{K} \leftarrow CandidateList(\bar{K}, \gamma)$;\\
		$k' \leftarrow Random(\hat{K})$;\\
		$y \leftarrow DijkstraLeader(\bar{f}^{k'},k'$);\\
		$\bar{K} \leftarrow \bar{K} \backslash \lbrace k' \rbrace$;\\
		
	}

	\For{$k \in K$}{
		$x \leftarrow DijkstraFollower(c,k)$;\\
	}
	$s \leftarrow \langle y,x \rangle$;\\
	$CloseEdge(s$);\\
	\If{$Cost(s) < MinCost$}{
	    $s_{best} \leftarrow s$;\\
	    $MinCost \leftarrow Cost(s_{best})$;\\
	}
	$\alpha \leftarrow \alpha - \frac{1}{MaxIterDP}$;\\
	$\bar{K} \leftarrow K$, $x \leftarrow 0$, $y \leftarrow 0$;\\
  }
  \Return{$s_{best}$}
}

\caption{\label{alg1}Partial Decoupling Heuristic}

\end{algorithm}

\noindent To solve the shortest path problem, the partial decoupling heuristic applies the Dijkstra algorithm. At the $|K|$ runs, the function \textit{DijkstraLeader} solves the problem of network construction, then, the shortest path problem is solved using the \textit {DijkstraFollower} function, generating a feasible solution. The notation $s \leftarrow \langle y,x \rangle$ means that the solution $s$ is storing the values of the variables $y$ and $x$ that were just defined by \textit{DijkstraLeader} and \textit{DijkstraFollower}. Since the function \textit{DijskstraLeader} can open edges that at the end do not have flow, we used the function \textit{CloseEdge}() set $y_{e} =0$ for every $x_{ij}^k =0 \mbox{, } \forall k \in K$. The $Random()$ function returns a random element from the set passed as a parameter. In order to choose the insertion order of the $|K|$ commodities, a candidate list consisting of a subset of commodities not yet routed, whose amount is greater than or equal to $ \gamma \%$ times the largest amount ($q_{k}$)  of the commodities not routed is create through the use of the function $CandidateList()$.

\subsection{LBound Method}
LBound Method is a strategy to probably find a stronger lower bound to the original problem. In order to do that, the method consists in relaxing all variables and at each iteration a subset of $y$ variables are turn into binary variables of the model (\ref{EQF}) - (\ref{FM}). The process repeats until $\lceil 0.2|E| \rceil$ iterations are done or an integer solution has been found. The number of iterations was decided after numerical experiments.
Details of the method could be seen in Algorithm \ref{alg:LBound}:

\begin{algorithm}[ht] 
\SetKwInOut{Input}{Input}
\Input{$K$, $\mathcal{G}$}
\KwData{$nvbin,cont \leftarrow 0$}
\Begin{
	$s_{inf} \leftarrow LinearRelaxion()$\;
	$\bar{E} \leftarrow E$\;
	\Repeat{$cont \geq \lceil 0.2|E| \rceil$ or $OptFound(s_{inf} )=TRUE$ or $nvbin > 0.9|E|$}{
			\For{$e \in \bar{E}$}{
				\If{$y_{e} \geq 0.5$}{
					$y_{e} \in  \{ 0,1 \} $\;
					$\bar{E} \setminus \{ e \}$\;
					$nvbin \leftarrow nvbin + 1$\;
				}
			}
			$s_{inf} \leftarrow SolveR()$\;
		
		$cont \leftarrow cont+1$\;
	}
	\Return{$s_{inf}$}
}

\caption{LBound}
\label{alg:LBound}
\end{algorithm}
\noindent The function $LinearRelaxation()$ solves the linear relaxation of the problem and returns the solution value. The function $SolveR()$ solves a relaxation the problem with a subset of binary variables. Function $OptFound()$ verifies if the solution found by the method is integer or not. It is important to remark that the condition $nvbin>0.9|E|$ was never reached.

\subsection{Variable Fixing Heuristic}
The Variable Fixing Heuristic (VFH) start using both the Partial Decoupling Heuristic and the LBound method. After applying those two methods, the VHF uses a relax and fix strategy to try to find a better solution.
Based on the Relax and Fix Heuristic \cite{Wolsey1998}, in this third part, we separate the variables in two distinct sets. $N_1$ is the set of relaxed variables and $N_2$ is the set of binary variables. Initialy $N_1$ contains all variables, while $N_2$ is empty. The main idea is at each iteration move a subset of the flow variables ($x^{k}$) from $N_1$ to $N_2$. At the end of each iteration, if a feasible solution for the relaxed model was found, the variables $y$ that are both zero and attend to the reduced cost criterion for variable fixing, are fixed as zero. The method repeats until all $x^{k}$ have been moved from $N_1$ to $N_2$ or the duality gap becomes lower than one.\\
In order to choose the order of $x^k$ variables to become binary, the procedure uses a candidate list. To choose a commodity, an element is randomly selected from a candidate list consisting of the commodities whose amount to be transported are greater than or equal to $\gamma \%$ times the largest amount of the commodity whose variables are not set as binary. A pseudo-code of the method is presented in Algorithm \ref{alg:VHF}.

\begin{algorithm}[htbp]
\SetKwInOut{Input}{Input}
\Input{$\gamma$, $K$, $\mathcal{G}$}
\KwData{$MinCost \leftarrow \infty$}

\Begin{
	$s_{best} \leftarrow \mbox{PartialDecoupling}(\gamma,K,\mathcal{G})$\;
	$s_{inf} \leftarrow LBound(K,\mathcal{G})$\;
	$MinCost \leftarrow Cost(s_{best})$ \;
	$\bar{K} \leftarrow K$\;
	\If{$OptFound(s_{inf} ) \neq TRUE$}{
	\While{$\bar{K} \not= \emptyset$ and $|s_{best}-s_{inf}| \geq 1$ }{
		$k \leftarrow CandidateList(\bar{K}, \gamma)$\;
		$x^k \in  \{ 0,1 \} $\;
		
		 $s \leftarrow SolveR(MinCost)$\;
		 \If{A feasible solution for the relaxed model was found}{
		 \For{$e \in E$}{
				\If{$y_e = 0$ and $RCVF(y_{e}) = TRUE$}{
					$y_e \leftarrow 0$\;
				}
			}
			
		\If{$Cost(s)<MinCost$ and $Feas(s)=TRUE$}{
			$s_{best} \leftarrow s$ \;
			$MinCost \leftarrow Cost(s_{best})$ \;	
		}
		\ElseIf{$Cost(s)>Cost(s_{inf})$ and $Feas(s)=FALSE$}{
			$s_{inf} \leftarrow s$\;	
			}
		}
		\Else{Exit}
		$\bar{K} \leftarrow \bar{K} \setminus \{ k \}$
	}
	\Return{$s_{best}$}
	}
	\Else{\Return{$s_{inf}$}}
}
\caption{VFH}
\label{alg:VHF}
\end{algorithm}
\pagebreak

\noindent The function $SolveR()$ solves a relaxation of the one level formulation (\ref{EQF})-(\ref{FM}) with a subset of binary variables, taking into consideration the primal bound $MinCost$. $MinCost$ is defined as the current best solution cost. The $RCVF()$ function returns TRUE if the Linear Relaxation cost plus the Reduced Cost of $y_{e}$ is greater than the current VFH solution. The function $Feas()$ returns true if the solution $s$ passed as parameter is a feasible solution to the original problem and returns false otherwise.\\

\subsection{Local Branching}
Introduced by Fiscetti and Lodi \cite{Fischetti2003}, the Local Branching (LB) technique could be used as a way of improving a given feasible solution. The LB makes use of a MIP solver to explore the solution subspaces effectively. The procedure can be seen as local search, but the neighborhoods are obtained through the introduction of linear inequalities in the MIP model, called local branching cuts. More specifically, the LB searches for a local optimum by restricting the number of variables, from the feasible solution, whose values can be changed.\\
Formally speaking, consider a feasible solution of the FCNDP-UOP, $s= \langle \bar{y} , \bar{x} \rangle  \in P$, where $P$ is the polyhedron formed by (\ref{EQF})-(\ref{FM}). The general idea would be adding the LB constraint
\begin{equation}
	\displaystyle\sum\limits_{e \in E | \bar{y}_{e} =0}y_{e} + \sum\limits_{e \in E | \bar{y}_{e} =1}(1 - y_{e}) \leq \Delta ,
\end{equation}

\noindent where $\Delta$ is a given positive integer parameter, indicating the number of variables $y_{e}$, $e \in E$, that are allowed to flip from one to zero and vice versa.\\
The strategy used here consists on applying the LB constraint only on $y$ variables, leaving  $x^{k}$ variables free of LB constraints.

\subsection{Ejection Cycle}
To understand the principles below the pertubation presented here, it is necessary to get to know a few metrics, developed by \cite{Paraskevopoulos2013}, to evaluate chains in a solution. \\
Consider a solution defined by the variables $x_{ij}^k$ for each arc $a \in A^E$ and each commodity $k \in K$ and  $y_{e}$ for each edge $e \in E$. For each open edge $e$, where $y_{e}=1$ and $x_{ij}^k > 0$ or $x_{ji}^k > 0$ for at least one commodity $k$, the edge \emph{inefficiency ratio} can be defined as:

\begin{equation}
	I_{e} = \frac{\displaystyle\sum_{k \in K}g_{ij}(x_{ij}^k+x_{ji}^k) + f_{e}}{\displaystyle\sum_{k \in K}(x_{ij}^k+x_{ji}^k)};\qquad \forall e=[i,j] \in E.
\end{equation}

\noindent The lower the value of $I_{e}$, more interesting it is to have edge $e$ in the solution. The average inefficiency ratio is defined as:

\begin{equation}
	\bar{I} = \frac{\displaystyle\sum_{e \in E}I_{e}y_{e}}{\displaystyle\sum_{e \in E}y_{e}}.
\end{equation}

\noindent With these metrics we can define a set of \emph{inefficient edges} as:

\begin{equation}
	A_{I} = \{ e \mbox{ }| \mbox{ } y_{e} = 1, I_{e} > \bar{I}  \} .
\end{equation}

\noindent As it can be seen above, the set of inefficient edges contains every edge in the solution whose inefficiency ratio is greater than the average inefficiency ratio. Our aim is to create a movement that remove flows from some of the inefficient edges in set $A_{I}$.\\
After evaluating the edges it is possible to construct \emph{inefficient chains} from a subset of the \emph{inefficient edges}. First, an edge is randomly chosen from the set $A_{I}$ of inefficient edges to form a component of the inefficient chain. If the current partial inefficient chain extends from node i to node j, then an edge $(a, i) \in A_{I}$ or $(j, b) \in A_{I}$ is added to the current chain, where nodes $a$ and $b$ are not included in the current chain. Whenever an edge is added to a chain, it is deleted from $A_{I}$. The process of extending the current chain continues until no further extension is possible or until the chain is composed by four edges. Unless $A_{I}$ is empty or contains a single arc, the process iterates with a random edge chosen to start a new chain. When the process ends, any chains containing a single edge is deleted. This is done in order to decrease the number of edges affected at each iteration of the method.\\
After constructing a set of \emph{inefficient chains}, we define our movement. The movement is defined analyzing each chain in the set of \emph{inefficient chains}.\\
The key aspect of our pertubation is the re-routing of flow from edges of the \emph{inefficient chain} to other edges of the network. First, a list of commodities ($K_{SET}$) that have a positive flow through at least one edge of the randomly selected \emph{inefficient chain} is formed. After that, the opening cost ($f_{e}$) of each edge in the \emph{inefficient chain} is set as infinity. After reassigning the costs, every commodity in $K_{SET}$ has its route destroyed and reconstructed by the Partial Decoupling Heuristic taking into account the new opening costs. If a feasible solution is found the method stops, else, another \emph{inefficient chain} is randomly selected and the process restarts. Algorithm \ref{alg:PFR} describes our Ejection Cycle procedure.

\begin{algorithm}[ht]
\SetKwInOut{Input}{Input}
\Input{$s$, $\gamma$, $K$, $\mathcal{G}$}

\Begin{
	
	$\mbox{P} \leftarrow PInefChain(s)$\;
	$\bar{s} \leftarrow \emptyset$ \;
	\While{$\mbox{P} \not= \emptyset \mbox{ and } \bar{s} \mbox{ is not feasible}$}{
		$rchain \leftarrow Random(\mbox{P})$\;
		$P \setminus \{rchain \}$\;
		$K_{SET} \leftarrow SK(s,rchain)$\;
		$\bar{s} \leftarrow \mbox{PartialDecoupling}(\mathcal{G},\gamma, K_{SET})$\;
		\If{$Cost(\bar{s}) \leq Cost(s)$}{
			$s \leftarrow \bar{s}$\;
			}
	}
	\Return{$s$}
}

\caption{Ejection Cycle}
 \label{alg:PFR}
\end{algorithm}
\pagebreak
\noindent In order to clarify Algorithm \ref{alg:PFR} it is necessary to define a few things. The function $PInefChain()$ returns the set $A_{I}$ of \emph{inefficient chains} in a solution $s$. The function $SK()$ returns the commodities that have a positive flow in solution $s$ through at least one arc of the \emph{inefficient chain} passed as parameter and set the fixed costs of the edges in the $rchain$ as infinity. The function PartialDecoupling() reroutes the commodities in  $K_{SET}$. In order to do that the \textit{DijkstraLeader} is applied for all $k \in K_{SET}$ and \textit{DijkstraFollower} for all $k \in K$. To account those changes, now the method PartialDecoupling() needs to receive a second parameter which is the set o commodities used in \textit{DijkstraLeader}.Besides that a partial solution for all $ k \in K \setminus K_{SET}$ is also passed as a parameter. 

\subsection{Iterated Local Search}
Developed by Louren\c{c}o et al. \cite{Lourenco2010}, the Iterated Local Search (ILS) is a metaheuristic that applies a local search method repeatedly to a set of solutions obtained by perturbing previously visited local optimal solutions. The ILS presented here uses as its main components, the VFH, the Local Branching and the Ejection Cycle presented in the previously subsections. The methods are applied in a straightforward way. First we ran the VFH to get a feasible solution and a lower bound. Secondly we try to improve the quality of the previously found solution through applying the Local Branching and the Ejection Cycle. The algorithm is described in Algorithm \ref{alg:ILS}.\\

\begin{algorithm}[ht] 
\SetKwInOut{Input}{Input}
\Input{$\gamma$, $\Delta$, $K$, $\mathcal{G}$}
\Begin{
    $s,s_{inf} \leftarrow \mbox{VFH}(\mathcal{G},K,\gamma)$\;
	$s \leftarrow \mbox{LB}(s,\Delta)$\;
	$\mbox{UpdateBest}(s)$\;
	\If{$|cost(s_{best})-cost(s_{inf})| \geq 1$}{
	\While{\mbox{Stop Criterion}=false}{
	$s \leftarrow \mbox{EjectionCycle}(s,\gamma, K, \mathcal{G})$\;
	$s \leftarrow \mbox{LB}(s,\Delta)$\;
	$\mbox{UpdateBest}(s)$\;
	}}
	\Return{$s_{best}$}
}

\caption{VFHLB}
\label{alg:ILS}
\end{algorithm}

\noindent In the VFHLB, the initial solution and the lower bound are generated by the VFH method. Then, the function LB performs the Local Branching as a Local Search and the EjectionCycle performs a perturbation.
\section{COMPUTATIONAL RESULTS}

In this section we present computational results for the VHFLB presented in the previous section. \\
The algorithm was coded in Xpress Mosel using FICO Xpress Optimization Suite, on an Intel (R) Core TM i3 - 3250 CPU @ 3.5 GHz computer with 8GB of RAM. Computing times are reported in seconds. In order to test the performance of the presented heuristic, we used networks data obtained from Mauttone, Labb\'{e} and Figueiredo \cite{Mauttone2008}.\\
In order to calibrate the algorithms we use $60 \%$ of our data so parameters overfitting could be avoided and the following $Stop Criterion$, $\gamma$ and $\Delta$ values were tested: $Stop Criterion = \{10\mbox{ iterations}; 50 \mbox{ iterations}; 100 \mbox{ iterations} \}$, $\gamma = \{0.75,0.85,0.90 \}$ and $\Delta = \{ \lceil \frac{|E|}{4} \rceil,\lceil \frac{|E|}{3} \rceil, \lceil \frac{|E|}{2} \rceil \}$.After the tests the parameters were calibrated as: $Stop Criterion = 10\mbox{ iterations}$, $\gamma=0.85$ and $\Delta= \lceil \frac{|E|}{2} \rceil$.\\
The data used are grouped according to the number of nodes in the graph (10, 20, 30), followed by the graph density (0.3, 0.5, 0.8) and finally the amount of different commodities to be transported (5, 10, 15, 20, 30, 45).\\
We are comparing the VFHLB results with the results of the GRASP presented by \cite{Gonzalez2013}, which, to the best of our knowledge, is the best heuristic aproach to solve the FCNDP-UOF. For the presented tables, we report the best solution (\emph{Best Sol}) and best time (\emph{Best Time}) reached by each  approach, the average gap (\emph{Avg GAP}) and the gap (\emph{GAP}) using the optimal solution. We also reported the average values for time (\emph{Avg Time}) and for solutions (\emph{Avg Sol}). Finally, it is reported standard deviation values for time (\emph{Dev Time}) and solution (\emph{Dev Sol}). The results in bold represent that the optimum has been found.

\begin{landscape}
\setlength{\tabcolsep}{3pt}
\begin{table*}[htbp]
\scriptsize
\centering
\begin{center}

 \resizebox{22cm}{!}{
\begin{tabular}{ccccccccccccccccc}

\hline
 & \multicolumn{8}{c}{GRASP} & \multicolumn{7}{c}{VFHLB} \\
\cmidrule(l){2-9} \cmidrule(rl){10-16}
& \textbf{Avg Sol} & \textbf{Avg Time} & \textbf{Dev Sol}  & \textbf{Dev Time} & \textbf{Best Sol} & \textbf{Best Time}& \textbf{Avg GAP} & \textbf{GAP} & \textbf{Avg Sol} & \textbf{Avg Time} &  \textbf{Dev Time} & \textbf{Best Sol} & \textbf{Best Time}& \textbf{Avg GAP} & \textbf{GAP} \\
\hline

\textbf{10-0.3-5-1} & \textbf{3942.00} & 1.2870 & 0.0000 & 0.0329 & \textbf{3942} & 1.2561 & 0.0000 & 0.0000 & \textbf{3942} & 0.0070 & 0.0017 & \textbf{3942} & 0.0060 & 0.0000 & 0.0000 \\ 
\textbf{10-0.3-5-2} & \textbf{4552.00} & 1.3267 & 0.0000 & 0.0172 & \textbf{4552} & 1.3110 & 0.0000 & 0.0000 & \textbf{4552} & 0.0038 & 0.0004 & \textbf{4552} & 0.0030 & 0.0000 & 0.0000 \\ 
\textbf{10-0.3-5-3} & \textbf{5762.00} & 1.2470 & 0.0000 & 0.0276 & \textbf{5762} & 1.2420 & 0.0000 & 0.0000 & \textbf{5762} & 0.0040 & 0.0000 & \textbf{5762} & 0.0040 & 0.0000 & 0.0000 \\ 
\textbf{10-0.3-5-4} & \textbf{4811.00} & 1.3150 & 0.0000 & 0.0230 & \textbf{4811} & 1.2834 & 0.0000 & 0.0000 & \textbf{4811} & 0.0044 & 0.0009 & \textbf{4811} & 0.0040 & 0.0000 & 0.0000 \\ 
\textbf{10-0.3-5-5} & \textbf{4831.00} & 1.3158 & 0.0000 & 0.0418 & \textbf{4831} & 1.3080 & 0.0000 & 0.0000 & \textbf{4831} & 0.0034 & 0.0005 & \textbf{4831} & 0.0030 & 0.0000 & 0.0000 \\ 
\textbf{10-0.3-10-1} & \textbf{8331.00} & 2.6486 & 0.0000 & 0.0462 & \textbf{8331} & 2.6380 & 0.0000 & 0.0000 & \textbf{8331} & 0.0136 & 0.0021 & \textbf{8331} & 0.0120 & 0.0000 & 0.0000 \\ 
\textbf{10-0.3-10-2} & \textbf{8812.00} & 2.8110 & 0.0000 & 0.0755 & \textbf{8812} & 2.7941 & 0.0000 & 0.0000 & \textbf{8812} & 0.0128 & 0.0024 & \textbf{8812} & 0.0110 & 0.0000 & 0.0000 \\ 
\textbf{10-0.3-10-3} & \textbf{10016.00} & 2.7410 & 0.0000 & 0.0395 & \textbf{10016} & 2.7246 & 0.0000 & 0.0000 & \textbf{10016} & 0.0080 & 0.0007 & \textbf{10016} & 0.0070 & 0.0000 & 0.0000 \\ 
\textbf{10-0.3-10-4} & \textbf{8750.00} & 2.6676 & 0.0000 & 0.0804 & \textbf{8750} & 2.6000 & 0.0000 & 0.0000 & \textbf{8750} & 0.0072 & 0.0004 & \textbf{8750} & 0.0070 & 0.0000 & 0.0000 \\ 
\textbf{10-0.3-10-5} & \textbf{10130.00} & 2.7004 & 0.0000 & 0.0847 & \textbf{10130} & 2.6950 & 0.0000 & 0.0000 & \textbf{10130} & 0.0186 & 0.0040 & \textbf{10130} & 0.0160 & 0.0000 & 0.0000 \\ 
\textbf{10-0.3-15-1} & \textbf{12490.00} & 4.1740 & 0.0000 & 0.1084 & \textbf{12490} & 4.1657 & 0.0000 & 0.0000 & \textbf{12490} & 0.0186 & 0.0036 & \textbf{12490} & 0.0170 & 0.0000 & 0.0000 \\ 
\textbf{10-0.3-15-2} & \textbf{17417.00} & 4.1920 & 0.0000 & 0.0762 & \textbf{17417} & 4.0662 & 0.0000 & 0.0000 & \textbf{17417} & 0.0208 & 0.0013 & \textbf{17417} & 0.0200 & 0.0000 & 0.0000 \\ 
\textbf{10-0.3-15-3} & \textbf{12378.00} & 4.2074 & 0.0000 & 0.1048 & \textbf{12378} & 4.1990 & 0.0000 & 0.0000 & \textbf{12378} & 0.0182 & 0.0045 & \textbf{12378} & 0.0150 & 0.0000 & 0.0000 \\ 
\textbf{10-0.3-15-4} & 11007.00 & 4.2281 & 0.0000 & 0.0549 & 11007 & 4.1210 & 0.0017 & 0.0017 & \textbf{10988} & 0.0196 & 0.0029 & \textbf{10988} & 0.0170 & 0.0000 & 0.0000 \\ 
\textbf{10-0.3-15-5} & \textbf{9066.00} & 4.2565 & 0.0000 & 0.0537 & \textbf{9066} & 4.2060 & 0.0000 & 0.0000 & \textbf{9066} & 0.0158 & 0.0008 & \textbf{9066} & 0.0150 & 0.0000 & 0.0000 \\ 
\textbf{20-0.3-10-1} & 6513.58 & 15.6530 & 136.4805 & 0.3393 & 6411 & 15.4965 & 0.0896 & 0.0724 & \textbf{5978} & 0.6980 & 0.0098 & \textbf{5978} & 0.6840 & 0.0000 & 0.0000 \\ 
\textbf{20-0.3-10-2} & 10813.30 & 16.5735 & 185.6884 & 0.5755 & 10664 & 16.3770 & 0.0329 & 0.0186 & \textbf{10469} & 4.7662 & 0.0886 & \textbf{10469} & 4.6650 & 0.0000 & 0.0000 \\ 
\textbf{20-0.3-10-3} & 7286.40 & 15.9854 & 132.1352 & 0.3434 & 7200 & 15.6720 & 0.0379 & 0.0256 & \textbf{7020} & 3.7044 & 0.1155 & \textbf{7020} & 3.5470 & 0.0000 & 0.0000 \\ 
\textbf{20-0.3-10-4} & 5754.74 & 15.8370 & 116.7287 & 0.3310 & 5598 & 15.7103 & 0.0494 & 0.0208 & \textbf{5484} & 2.7238 & 0.0806 & \textbf{5484} & 2.6230 & 0.0000 & 0.0000 \\ 
\textbf{20-0.3-10-5} & 8322.00 & 16.0420 & 0.0000 & 0.3995 & 8322 & 16.0100 & 0.0492 & 0.0492 & \textbf{7932} & 14.4424 & 0.2933 & \textbf{7932} & 14.1280 & 0.0000 & 0.0000 \\ 
\textbf{20-0.3-20-1} & \textbf{9488.00} & 32.0957 & 0.0000 & 1.3602 & \textbf{9488} & 31.8410 & 0.0000 & 0.0000 & \textbf{9488} & 0.8662 & 0.0272 & \textbf{9488} & 0.8400 & 0.0000 & 0.0000 \\ 
\textbf{20-0.3-20-2} & 11699.86 & 31.6390 & 201.3070 & 0.9075 & 11607 & 30.9429 & 0.0155 & 0.0075 & \textbf{11521} & 3.3546 & 0.1505 & \textbf{11521} & 3.2080 & 0.0000 & 0.0000 \\ 
\textbf{20-0.3-20-3} & 8670.82 & 32.5660 & 222.8998 & 0.7159 & 8568 & 32.4357 & 0.0485 & 0.0360 & \textbf{8270} & 1.2644 & 0.0393 & \textbf{8270} & 1.2280 & 0.0000 & 0.0000 \\ 
\textbf{20-0.3-20-4} & 12320.58 & 31.9430 & 300.0561 & 1.0738 & 11985 & 31.6236 & 0.0353 & 0.0071 & \textbf{11901} & 21.8506 & 0.9442 & \textbf{11901} & 21.0000 & 0.0000 & 0.0000 \\ 
\textbf{20-0.3-20-5} & 10379.38 & 32.1230 & 178.5869 & 0.4624 & 10297 & 31.9303 & 0.0749 & 0.0664 & \textbf{9656} & 1.8926 & 0.0947 & \textbf{9656} & 1.8190 & 0.0000 & 0.0000 \\ 
\textbf{20-0.3-30-1} & 13244.00 & 49.2763 & 0.0000 & 0.7556 & 13244 & 48.6920 & 0.0587 & 0.0587 & \textbf{12510} & 1.4656 & 0.0292 & \textbf{12510} & 1.4280 & 0.0000 & 0.0000 \\ 
\textbf{20-0.3-30-2} & 14854.90 & 49.8060 & 364.8115 & 1.7615 & 14737 & 49.4076 & 0.0449 & 0.0366 & \textbf{14216} & 2.2224 & 0.1063 & \textbf{14216} & 2.1130 & 0.0000 & 0.0000 \\ 
\textbf{20-0.3-30-3} & 14687.52 & 48.1790 & 577.2804 & 1.4053 & 14629 & 47.7936 & 0.0967 & 0.0923 & \textbf{13393} & 5.2596 & 0.1448 & \textbf{13393} & 5.0720 & 0.0000 & 0.0000 \\ 
\textbf{20-0.3-30-4} & 15420.97 & 48.6160 & 327.7683 & 0.6324 & 15329 & 48.3243 & 0.0670 & 0.0607 & \textbf{14452} & 1.7608 & 0.0733 & \textbf{14452} & 1.6980 & 0.0000 & 0.0000 \\ 
\textbf{20-0.3-30-5} & 12599.00 & 51.3221 & 0.0000 & 1.0764 & 12599 & 51.0160 & 0.1033 & 0.1033 & \textbf{11419} & 1.3276 & 0.0398 & \textbf{11419} & 1.2950 & 0.0000 & 0.0000 \\ 
\textbf{30-0.3-15-1} & 8529.32 & 69.3908 & 263.2338 & 1.5946 & 8395 & 68.5680 & 0.0879 & 0.0708 & \textbf{7840} & 2.3482 & 0.0674 & \textbf{7840} & 2.2900 & 0.0000 & 0.0000 \\ 
\textbf{30-0.3-15-2} & 10051.33 & 65.7535 & 340.4006 & 1.0051 & 10112 & 64.7180 & 0.0604 & 0.0668 & \textbf{9479} & 11.9144 & 0.2141 & \textbf{9479} & 11.6160 & 0.0000 & 0.0000 \\ 
\textbf{30-0.3-15-3} & 7422.75 & 66.0270 & 196.0199 & 1.8967 & 7281 & 65.7629 & 0.0536 & 0.0335 & \textbf{7045} & 5.4786 & 0.0389 & \textbf{7045} & 5.4180 & 0.0000 & 0.0000 \\ 
\textbf{30-0.3-15-4} & 8775.16 & 66.4171 & 168.0749 & 2.3415 & 8654 & 65.8900 & 0.0414 & 0.0271 & \textbf{8426} & 26.4730 & 0.5365 & \textbf{8426} & 25.7670 & 0.0000 & 0.0000 \\ 
\textbf{30-0.3-15-5} & 9626.00 & 66.1244 & 0.0000 & 2.0463 & 9626 & 65.7300 & 0.0949 & 0.0949 & \textbf{8792} & 98.2168 & 1.1438 & \textbf{8792} & 97.3190 & 0.0000 & 0.0000 \\ 
\textbf{30-0.3-30-1} & 15766.28 & 133.4690 & 287.2792 & 2.8935 & 15286 & 132.1343 & 0.1927 & 0.1564 & \textbf{13219} & 9.4686 & 0.0500 & \textbf{13219} & 9.4110 & 0.0000 & 0.0000 \\ 
\textbf{30-0.3-30-2} & 14308.35 & 138.7550 & 252.7530 & 2.3416 & 13973 & 137.6450 & 0.0908 & 0.0653 & \textbf{13117} & 35.3648 & 0.6179 & \textbf{13117} & 34.8360 & 0.0000 & 0.0000 \\ 
\textbf{30-0.3-30-3} & 15504.47 & 139.7580 & 580.6050 & 3.8356 & 15412 & 137.8014 & 0.1450 & 0.1382 & \textbf{13541} & 18.5124 & 0.3369 & \textbf{13541} & 18.2120 & 0.0000 & 0.0000 \\ 
\textbf{30-0.3-30-4} & 14766.19 & 132.6110 & 254.0662 & 2.3091 & 14649 & 130.7544 & 0.1546 & 0.1454 & \textbf{12789} & 31.2224 & 1.5092 & \textbf{12789} & 29.8950 & 0.0000 & 0.0000 \\ 
\textbf{30-0.3-30-5} & 13841.41 & 133.6140 & 307.9978 & 3.7867 & 13517 & 133.0795 & 0.1634 & 0.1362 & \textbf{11897} & 8.3360 & 0.3231 & \textbf{11897} & 8.0590 & 0.0000 & 0.0000 \\ 
\textbf{30-0.3-45-1} & 18885.64 & 204.8792 & 663.5134 & 2.6948 & 18773 & 200.8620 & 0.1849 & 0.1779 & \textbf{15938} & 16.7946 & 0.6475 & \textbf{15938} & 16.4120 & 0.0000 & 0.0000 \\ 
\textbf{30-0.3-45-2} & 14455.60 & 206.9196 & 597.8858 & 4.3356 & 14200 & 203.6610 & 0.0955 & 0.0761 & \textbf{13196} & 29.0830 & 0.5499 & \textbf{13196} & 28.5450 & 0.0000 & 0.0000 \\ 
\textbf{30-0.3-45-3} & 19346.43 & 202.7890 & 340.7032 & 6.4728 & \textbf{18893} & 202.3834 & 0.0240 & 0.0000 & \textbf{18893} & 230.5346 & 8.5348 & \textbf{18893} & 223.8230 & 0.0000 & 0.0000 \\ 
\textbf{30-0.3-45-4} & 19162.29 & 215.2056 & 637.8094 & 4.2326 & 19048 & 209.7520 & 0.0870 & 0.0805 & \textbf{17629} & 29.6728 & 0.6131 & \textbf{17629} & 29.2020 & 0.0000 & 0.0000 \\ 
\textbf{30-0.3-45-5} & 17909.32 & 205.4560 & 231.1970 & 6.6092 & 17732 & 200.9360 & 0.0926 & 0.0817 & \textbf{16392} & 250.6620 & 4.4910 & \textbf{16392} & 246.4560 & 0.0000 & 0.0000 \\ 

 \hline

{\bf Avg } &    11171.1236	& 57.2432
     &            &            &    11078.3111	& 56.5236&	0.0528 &	0.0446&	10537.2889	&19.3746
&           & 10537.2889&	18.9504 & 0.0000 &	0.0000
\\
\hline

\end{tabular}  
 }
\end{center}
\caption{Computational results for GRASP and VFHLB approaches for $0.3$ density instances \label{tab01}}
\end{table*}

\end{landscape}

\begin{landscape}

\setlength{\tabcolsep}{3pt}
\begin{table*}[htbp]
\scriptsize
\centering
\begin{center}

 \resizebox{22cm}{!}{
\begin{tabular}{ccccccccccccccccc}

\hline
 & \multicolumn{8}{c}{GRASP} & \multicolumn{7}{c}{VFHLB} \\
\cmidrule(l){2-9} \cmidrule(rl){10-16}
& \textbf{Avg Sol} & \textbf{Avg Time} & \textbf{Dev Sol}  & \textbf{Dev Time} & \textbf{Best Sol} & \textbf{Best Time}& \textbf{Avg GAP} & \textbf{GAP} & \textbf{Avg Sol} & \textbf{Avg Time} &  \textbf{Dev Time} & \textbf{Best Sol} & \textbf{Best Time}& \textbf{Avg GAP} & \textbf{GAP} \\
\hline

\textbf{10-0.5-5-1} & \textbf{4360.00} & 1.8240 & 0.0000 & 0.0568 & \textbf{4360} & 1.8058 & 0.0000 & 0.0000 & \textbf{4360} & 0.0086 & 0.0005 & \textbf{4360} & 0.0080 & 0.0000 & 0.0000 \\ 
\textbf{10-0.5-5-2} & \textbf{1351.00} & 1.9186 & 0.0000 & 0.0632 & \textbf{1351} & 1.9110 & 0.0000 & 0.0000 & \textbf{1351} & 0.0092 & 0.0004 & \textbf{1351} & 0.0090 & 0.0000 & 0.0000 \\ 
\textbf{10-0.5-5-3} & \textbf{2932.00} & 1.8339 & 0.0000 & 0.0533 & \textbf{2932} & 1.8050 & 0.0000 & 0.0000 & \textbf{2932} & 0.0082 & 0.0011 & \textbf{2932} & 0.0070 & 0.0000 & 0.0000 \\ 
\textbf{10-0.5-5-4} & \textbf{4920.00} & 1.9230 & 0.0000 & 0.0662 & \textbf{4920} & 1.8890 & 0.0000 & 0.0000 & \textbf{4920} & 0.2516 & 0.0138 & \textbf{4920} & 0.2430 & 0.0000 & 0.0000 \\ 
\textbf{10-0.5-5-5} & \textbf{4469.00} & 1.8880 & 0.0000 & 0.0604 & \textbf{4469} & 1.8730 & 0.0000 & 0.0000 & \textbf{4469} & 0.0184 & 0.0009 & \textbf{4469} & 0.0180 & 0.0000 & 0.0000 \\ 
\textbf{10-0.5-10-1} & 7566.00 & 3.7367 & 0.0040 & 0.1192 & 7566 & 3.7070 & 0.0040 & 0.0040 & \textbf{7536} & 0.0542 & 0.0008 & \textbf{7536} & 0.0530 & 0.0000 & 0.0000 \\ 
\textbf{10-0.5-10-2} & 7575.96 & 3.7589 & 193.3202 & 0.0645 & 7442 & 3.7070 & 0.0431 & 0.0246 & \textbf{7263} & 0.3026 & 0.0027 & \textbf{7263} & 0.3000 & 0.0000 & 0.0000 \\ 
\textbf{10-0.5-10-3} & 5399.55 & 3.7424 & 131.8461 & 0.0968 & \textbf{5273} & 3.6980 & 0.0240 & 0.0000 & \textbf{5273} & 0.0166 & 0.0005 & \textbf{5273} & 0.0160 & 0.0000 & 0.0000 \\ 
\textbf{10-0.5-10-4} & 5983.61 & 3.8770 & 105.7580 & 0.0847 & 5901 & 3.8460 & 0.0221 & 0.0080 & \textbf{5854} & 0.0174 & 0.0009 & \textbf{5854} & 0.0170 & 0.0000 & 0.0000 \\ 
\textbf{10-0.5-10-5} & 5102.45 & 3.7687 & 66.9842 & 0.0719 & 5032 & 3.7240 & 0.0240 & 0.0098 & \textbf{4983} & 0.8284 & 0.0187 & \textbf{4983} & 0.8060 & 0.0000 & 0.0000 \\ 
\textbf{10-0.5-15-1} & 9379.00 & 5.6480 & 0.0041 & 0.1157 & 9379 & 5.5350 & 0.0041 & 0.0041 & \textbf{9341} & 0.0312 & 0.0013 & \textbf{9341} & 0.0300 & 0.0000 & 0.0000 \\ 
\textbf{10-0.5-15-2} & 7512.00 & 5.7720 & 0.0000 & 0.0759 & 7512 & 5.7027 & 0.1264 & 0.1264 & \textbf{6669} & 0.0236 & 0.0013 & \textbf{6669} & 0.0220 & 0.0000 & 0.0000 \\ 
\textbf{10-0.5-15-3} & \textbf{10324.00} & 5.9603 & 0.0000 & 0.1085 & \textbf{10324} & 5.9130 & 0.0000 & 0.0000 & \textbf{10324} & 0.3338 & 0.0041 & \textbf{10324} & 0.3300 & 0.0000 & 0.0000 \\ 
\textbf{10-0.5-15-4} & \textbf{6339.00} & 5.9380 & 0.0000 & 0.2099 & \textbf{6339} & 5.8100 & 0.0000 & 0.0000 & \textbf{6339} & 0.0810 & 0.0025 & \textbf{6339} & 0.0790 & 0.0000 & 0.0000 \\ 
\textbf{10-0.5-15-5} & 9519.00 & 5.9964 & 0.0002 & 0.1417 & 9519 & 5.9370 & 0.0002 & 0.0002 & \textbf{9517} & 4.0846 & 0.0354 & \textbf{9517} & 4.0300 & 0.0000 & 0.0000 \\ 
\textbf{20-0.5-10-1} & \textbf{4784.00} & 21.5620 & 0.0000 & 0.8304 & \textbf{4784} & 21.4326 & 0.0000 & 0.0000 & \textbf{4784} & 2.6538 & 0.0199 & \textbf{4784} & 2.6310 & 0.0000 & 0.0000 \\ 
\textbf{20-0.5-10-2} & \textbf{7689.00} & 21.8640 & 0.0000 & 0.5656 & \textbf{7689} & 21.7328 & 0.0000 & 0.0000 & \textbf{7689} & 1.9200 & 0.0466 & \textbf{7689} & 1.8770 & 0.0000 & 0.0000 \\ 
\textbf{20-0.5-10-3} & \textbf{6184.00} & 22.6760 & 0.0000 & 0.4702 & \textbf{6184} & 22.4492 & 0.0000 & 0.0000 & \textbf{6184} & 0.5824 & 0.0102 & \textbf{6184} & 0.5670 & 0.0000 & 0.0000 \\ 
\textbf{20-0.5-10-4} & 5532.91 & 22.4149 & 95.1989 & 0.2894 & 5489 & 22.1930 & 0.0663 & 0.0578 & \textbf{5189} & 1.6642 & 0.0275 & \textbf{5189} & 1.6330 & 0.0000 & 0.0000 \\ 
\textbf{20-0.5-10-5} & 6233.72 & 22.7810 & 80.4730 & 0.5918 & 6172 & 22.7354 & 0.0302 & 0.0200 & \textbf{6051} & 26.7656 & 0.0977 & \textbf{6051} & 26.6630 & 0.0000 & 0.0000 \\ 
\textbf{20-0.5-20-1} & 9964.00 & 46.5030 & 0.0000 & 0.9544 & 9964 & 45.8520 & 0.1302 & 0.1302 & \textbf{8816} & 2.9528 & 0.0153 & \textbf{8816} & 2.9320 & 0.0000 & 0.0000 \\ 
\textbf{20-0.5-20-2} & 8721.34 & 47.4527 & 150.4528 & 1.8322 & \textbf{8584} & 46.8900 & 0.0160 & 0.0000 & \textbf{8584} & 4.4280 & 0.0511 & \textbf{8584} & 4.3720 & 0.0000 & 0.0000 \\ 
\textbf{20-0.5-20-3} & 8354.83 & 45.7165 & 214.8412 & 0.9228 & 8305 & 44.6450 & 0.1051 & 0.0985 & \textbf{7560} & 7.0656 & 0.0300 & \textbf{7560} & 7.0130 & 0.0000 & 0.0000 \\ 
\textbf{20-0.5-20-4} & 7750.74 & 45.2840 & 100.0567 & 0.8360 & 7674 & 44.9217 & 0.0153 & 0.0052 & \textbf{7634} & 1.5694 & 0.0201 & \textbf{7634} & 1.5470 & 0.0000 & 0.0000 \\ 
\textbf{20-0.5-20-5} & 8636.00 & 44.8590 & 0.0000 & 1.1159 & 8636 & 44.7693 & 0.0443 & 0.0443 & \textbf{8270} & 6.0790 & 0.0509 & \textbf{8270} & 6.0160 & 0.0000 & 0.0000 \\ 
\textbf{20-0.5-30-1} & 12600.00 & 67.9890 & 0.0000 & 2.3355 & 12600 & 67.9890 & 0.2406 & 0.2406 & \textbf{10156} & 1.8056 & 0.0785 & \textbf{10156} & 1.7300 & 0.0000 & 0.0000 \\ 
\textbf{20-0.5-30-2} & 12932.00 & 68.6630 & 0.0000 & 1.9053 & 12932 & 68.6630 & 0.1341 & 0.1341 & \textbf{11403} & 7.2198 & 0.2475 & \textbf{11403} & 7.0420 & 0.0000 & 0.0000 \\ 
\textbf{20-0.5-30-3} & 13021.40 & 73.2877 & 334.7399 & 1.3527 & 12867 & 71.5700 & 0.1225 & 0.1092 & \textbf{11600} & 13.7846 & 0.3707 & \textbf{11600} & 13.5040 & 0.0000 & 0.0000 \\ 
\textbf{20-0.5-30-4} & 12333.56 & 70.8795 & 317.1527 & 1.3237 & 12260 & 68.8150 & 0.0465 & 0.0403 & \textbf{11785} & 6.8018 & 0.0628 & \textbf{11785} & 6.7190 & 0.0000 & 0.0000 \\ 
\textbf{20-0.5-30-5} & 10989.00 & 69.4657 & 0.0000 & 1.8168 & 10989 & 69.3270 & 0.1496 & 0.1496 & \textbf{9559} & 7.3206 & 0.0511 & \textbf{9559} & 7.2530 & 0.0000 & 0.0000 \\ 
\textbf{30-0.5-15-1} & 6824.93 & 104.3949 & 112.2020 & 3.3325 & 6744 & 103.9790 & 0.1707 & 0.1568 & \textbf{5830} & 12.4814 & 0.1506 & \textbf{5830} & 12.3470 & 0.0000 & 0.0000 \\ 
\textbf{30-0.5-15-2} & 6888.00 & 103.6410 & 0.0000 & 4.0814 & 6888 & 102.8119 & 0.0527 & 0.0527 & \textbf{6543} & 20.0704 & 0.1470 & \textbf{6543} & 19.8560 & 0.0000 & 0.0000 \\ 
\textbf{30-0.5-15-3} & 5809.89 & 109.4442 & 52.8671 & 2.4582 & 5741 & 107.5090 & 0.0223 & 0.0102 & \textbf{5683} & 14.5294 & 0.1577 & \textbf{5683} & 14.3380 & 0.0000 & 0.0000 \\ 
\textbf{30-0.5-15-4} & 6097.00 & 106.1190 & 0.0000 & 2.2691 & 6097 & 103.3599 & 0.0919 & 0.0919 & \textbf{5584} & 11.6152 & 0.1212 & \textbf{5584} & 11.4330 & 0.0000 & 0.0000 \\ 
\textbf{30-0.5-15-5} & \textbf{5794.00} & 108.9972 & 0.0000 & 3.4635 & \textbf{5794} & 107.9180 & 0.0000 & 0.0000 & \textbf{5794} & 22.8150 & 1.1958 & \textbf{5794} & 21.9610 & 0.0000 & 0.0000 \\ 
\textbf{30-0.5-30-1} & 8823.02 & 209.1856 & 151.8084 & 6.4735 & 8753 & 207.9380 & 0.0274 & 0.0192 & \textbf{8588} & 28.3988 & 0.5552 & \textbf{8588} & 27.9600 & 0.0000 & 0.0000 \\ 
\textbf{30-0.5-30-2} & 9134.00 & 212.8090 & 0.0000 & 3.6315 & 9134 & 211.9578 & 0.0432 & 0.0432 & \textbf{8756} & 56.7008 & 1.7342 & \textbf{8756} & 55.3860 & 0.0000 & 0.0000 \\ 
\textbf{30-0.5-30-3} & 10908.73 & 206.0050 & 138.0897 & 4.4661 & \textbf{10591} & 203.9450 & 0.0300 & 0.0000 & \textbf{10591} & 445.2772 & 13.4245 & \textbf{10591} & 432.3970 & 0.0000 & 0.0000 \\ 
\textbf{30-0.5-30-4} & 9120.14 & 210.4039 & 227.4169 & 6.4708 & 9012 & 209.1490 & 0.1240 & 0.1107 & \textbf{8114} & 74.5170 & 1.2355 & \textbf{8114} & 73.2920 & 0.0000 & 0.0000 \\ 
\textbf{30-0.5-30-5} & 13575.00 & 214.7650 & 0.0000 & 3.1942 & 13575 & 209.1811 & 0.0700 & 0.0700 & \textbf{12687} & 117.1688 & 2.5543 & \textbf{12687} & 114.8630 & 0.0000 & 0.0000 \\ 
\textbf{30-0.5-45-1} & 11160.00 & 332.1730 & 0.0000 & 13.8050 & 11160 & 330.1800 & 0.0953 & 0.0953 & \textbf{10189} & 31.7414 & 0.3404 & \textbf{10189} & 31.3220 & 0.0000 & 0.0000 \\ 
\textbf{30-0.5-45-2} & 12105.07 & 319.6155 & 248.6762 & 7.4214 & 12009 & 316.4510 & 0.1540 & 0.1448 & \textbf{10490} & 67.9630 & 2.6552 & \textbf{10490} & 65.6430 & 0.0000 & 0.0000 \\ 
\textbf{30-0.5-45-3} & 15733.00 & 324.8540 & 0.0000 & 6.0844 & 15733 & 315.7581 & 0.1509 & 0.1509 & \textbf{13670} & 150.6902 & 8.3190 & \textbf{13670} & 142.3510 & 0.0000 & 0.0000 \\ 
\textbf{30-0.5-45-4} & 10910.00 & 322.2408 & 0.0000 & 4.1851 & 10910 & 316.5430 & 0.1321 & 0.1321 & \textbf{9637} & 76.2042 & 2.7176 & \textbf{9637} & 73.6180 & 0.0000 & 0.0000 \\ 
\textbf{30-0.5-45-5} & 12870.05 & 314.7070 & 267.3752 & 5.6402 & 12593 & 310.3011 & 0.1086 & 0.0848 & \textbf{11609} & 17.7640 & 0.4141 & \textbf{11609} & 17.3510 & 0.0000 & 0.0000 \\ 

  \hline

{\bf Avg } &    8315.8204 & 	87.7409     &            &            &       8270.7111&	86.6185&	0.0583	& 0.0527	&7781.3333	&27.7027&           & 7781.3333 &	26.9241 & 0.0000 &	0.0000
\\
\hline

\end{tabular}  
 }
\end{center}
\caption{Computational results for GRASP and VFHLB approaches for $0.5$ density instances \label{tab02}}
\end{table*}
\end{landscape}

\begin{landscape}

\setlength{\tabcolsep}{3pt}
\begin{table*}[htbp]
\scriptsize
\centering
\begin{center}

 \resizebox{22cm}{!}{
\begin{tabular}{ccccccccccccccccc}

\hline
 & \multicolumn{8}{c}{GRASP} & \multicolumn{7}{c}{VFHLB} \\
\cmidrule(l){2-9} \cmidrule(rl){10-16}
& \textbf{Avg Sol} & \textbf{Avg Time} & \textbf{Dev Sol}  & \textbf{Dev Time} & \textbf{Best Sol} & \textbf{Best Time}& \textbf{Avg GAP} & \textbf{GAP} & \textbf{Avg Sol} & \textbf{Avg Time} &  \textbf{Dev Time} & \textbf{Best Sol} & \textbf{Best Time}& \textbf{Avg GAP} & \textbf{GAP} \\
\hline

\textbf{10-0.8-5-1} & 4033.83 & 3.0370 & 108.4895 & 0.0314 & 3986 & 3.0249 & 0.1146 & 0.1014 & \textbf{3619} & 0.0142 & 0.0004 & \textbf{3619} & 0.0140 & 0.0000 & 0.0000 \\ 
\textbf{10-0.8-5-2} & 3535.68 & 2.9300 & 60.9944 & 0.0883 & \textbf{3480} & 2.9100 & 0.0160 & 0.0000 & \textbf{3480} & 0.1480 & 0.0016 & \textbf{3480} & 0.1460 & 0.0000 & 0.0000 \\ 
\textbf{10-0.8-5-3} & 3330.27 & 2.7480 & 112.2499 & 0.0442 & 3317 & 2.7150 & 0.1035 & 0.0991 & \textbf{3018} & 0.2422 & 0.0041 & \textbf{3018} & 0.2380 & 0.0000 & 0.0000 \\ 
\textbf{10-0.8-5-4} & \textbf{3518.00} & 2.9770 & 0.0000 & 0.0614 & \textbf{3518} & 2.8758 & 0.0000 & 0.0000 & \textbf{3518} & 0.0886 & 0.0009 & \textbf{3518} & 0.0880 & 0.0000 & 0.0000 \\ 
\textbf{10-0.8-5-5} & 3960.68 & 2.7730 & 80.1635 & 0.0521 & 3906 & 2.7620 & 0.0232 & 0.0090 & \textbf{3871} & 0.0390 & 0.0012 & \textbf{3871} & 0.0380 & 0.0000 & 0.0000 \\ 
\textbf{10-0.8-10-1} & 6031.84 & 5.9723 & 114.2613 & 0.0866 & 5902 & 5.8210 & 0.0376 & 0.0153 & \textbf{5813} & 0.0458 & 0.0046 & \textbf{5813} & 0.0430 & 0.0000 & 0.0000 \\ 
\textbf{10-0.8-10-2} & 5120.64 & 5.7880 & 107.2940 & 0.1479 & \textbf{5040} & 5.6954 & 0.0160 & 0.0000 & \textbf{5040} & 1.0682 & 0.0285 & \textbf{5040} & 1.0440 & 0.0000 & 0.0000 \\ 
\textbf{10-0.8-10-3} & 3975.00 & 5.9039 & 0.0000 & 0.1344 & 3975 & 5.8570 & 0.1360 & 0.1360 & \textbf{3499} & 0.0826 & 0.0018 & \textbf{3499} & 0.0810 & 0.0000 & 0.0000 \\ 
\textbf{10-0.8-10-4} & 5460.55 & 5.9090 & 116.8811 & 0.1723 & \textbf{5364} & 5.7908 & 0.0180 & 0.0000 & \textbf{5364} & 0.0876 & 0.0086 & \textbf{5364} & 0.0830 & 0.0000 & 0.0000 \\ 
\textbf{10-0.8-10-5} & 4225.54 & 5.7690 & 72.7043 & 0.1662 & 4192 & 5.7690 & 0.0224 & 0.0143 & \textbf{4133} & 1.1376 & 0.0334 & \textbf{4133} & 1.0860 & 0.0000 & 0.0000 \\ 
\textbf{10-0.8-15-1} & 6976.61 & 8.9230 & 90.2083 & 0.3180 & 6935 & 8.8338 & 0.0227 & 0.0166 & \textbf{6822} & 0.1478 & 0.0040 & \textbf{6822} & 0.1440 & 0.0000 & 0.0000 \\ 
\textbf{10-0.8-15-2} & 5276.29 & 8.8852 & 77.3640 & 0.1640 & \textbf{5183} & 8.6770 & 0.0180 & 0.0000 & \textbf{5183} & 0.1492 & 0.0027 & \textbf{5183} & 0.1450 & 0.0000 & 0.0000 \\ 
\textbf{10-0.8-15-3} & 5017.00 & 9.0300 & 0.0000 & 0.0780 & 5017 & 8.9940 & 0.1092 & 0.1092 & \textbf{4523} & 0.6238 & 0.0080 & \textbf{4523} & 0.6140 & 0.0000 & 0.0000 \\ 
\textbf{10-0.8-15-4} & 7663.62 & 8.9097 & 64.8997 & 0.2998 & \textbf{7484} & 8.8390 & 0.0240 & 0.0000 & \textbf{7484} & 0.6206 & 0.0086 & \textbf{7484} & 0.6070 & 0.0000 & 0.0000 \\ 
\textbf{10-0.8-15-5} & 4751.60 & 9.2254 & 85.5372 & 0.2468 & 4686 & 9.2070 & 0.2364 & 0.2194 & \textbf{3843} & 0.4682 & 0.0066 & \textbf{3843} & 0.4610 & 0.0000 & 0.0000 \\ 
\textbf{20-0.8-10-1} & 4120.80 & 34.3230 & 105.3503 & 0.8950 & 4040 & 34.3230 & 0.0440 & 0.0236 & \textbf{3947} & 0.6520 & 0.0107 & \textbf{3947} & 0.6440 & 0.0000 & 0.0000 \\ 
\textbf{20-0.8-10-2} & 3915.00 & 34.5080 & 0.0000 & 1.1326 & 3915 & 34.0249 & 0.0460 & 0.0460 & \textbf{3743} & 6.9310 & 0.2826 & \textbf{3743} & 6.7250 & 0.0000 & 0.0000 \\ 
\textbf{20-0.8-10-3} & 3480.24 & 34.8060 & 74.7532 & 0.5791 & \textbf{3412} & 34.3883 & 0.0200 & 0.0000 & \textbf{3412} & 0.1918 & 0.0033 & \textbf{3412} & 0.1880 & 0.0000 & 0.0000 \\ 
\textbf{20-0.8-10-4} & 4209.00 & 35.2740 & 0.0000 & 0.8032 & 4209 & 34.9940 & 0.0301 & 0.0301 & \textbf{4086} & 5.0812 & 0.1772 & \textbf{4086} & 4.9450 & 0.0000 & 0.0000 \\ 
\textbf{20-0.8-10-5} & 4542.98 & 35.6360 & 97.5143 & 0.7726 & \textbf{4498} & 35.2796 & 0.0100 & 0.0000 & \textbf{4498} & 4.6612 & 0.0678 & \textbf{4498} & 4.6030 & 0.0000 & 0.0000 \\ 
\textbf{20-0.8-20-1} & 6909.00 & 70.8823 & 0.0000 & 1.7308 & 6909 & 69.2210 & 0.1920 & 0.1920 & \textbf{5796} & 4.2190 & 0.0869 & \textbf{5796} & 4.1280 & 0.0000 & 0.0000 \\ 
\textbf{20-0.8-20-2} & 7635.54 & 71.4810 & 187.0284 & 1.0189 & 7590 & 70.3373 & 0.0851 & 0.0786 & \textbf{7037} & 313.3302 & 20.9517 & \textbf{7037} & 297.9690 & 0.0000 & 0.0000 \\ 
\textbf{20-0.8-20-3} & 6251.89 & 68.9992 & 89.4775 & 1.8381 & 5422 & 68.1810 & 0.3603 & 0.1797 & \textbf{4596} & 5.2952 & 0.1021 & \textbf{4596} & 5.2230 & 0.0000 & 0.0000 \\ 
\textbf{20-0.8-20-4} & 5187.00 & 70.2559 & 69.0130 & 2.4494 & 5250 & 69.9760 & 0.0693 & 0.0823 & \textbf{4851} & 2.8762 & 0.0466 & \textbf{4851} & 2.8170 & 0.0000 & 0.0000 \\ 
\textbf{20-0.8-20-5} & 6855.53 & 72.1322 & 86.2333 & 1.9296 & 6267 & 71.4180 & 0.1264 & 0.0297 & \textbf{6086} & 10.8284 & 0.5098 & \textbf{6086} & 10.5270 & 0.0000 & 0.0000 \\ 
\textbf{20-0.8-30-1} & 9425.00 & 105.0060 & 0.0000 & 2.1653 & 9425 & 101.2258 & 0.2132 & 0.2132 & \textbf{7769} & 7.7738 & 0.0747 & \textbf{7769} & 7.7040 & 0.0000 & 0.0000 \\ 
\textbf{20-0.8-30-2} & 8735.33 & 110.7691 & 126.4167 & 1.9805 & 8666 & 109.8900 & 0.1373 & 0.1282 & \textbf{7681} & 14.1722 & 0.2527 & \textbf{7681} & 13.9840 & 0.0000 & 0.0000 \\ 
\textbf{20-0.8-30-3} & 5947.89 & 107.2994 & 201.4348 & 2.6665 & 5889 & 106.2370 & 0.1563 & 0.1448 & \textbf{5144} & 14.6920 & 0.3542 & \textbf{5144} & 14.4420 & 0.0000 & 0.0000 \\ 
\textbf{20-0.8-30-4} & 8768.08 & 104.7711 & 177.5349 & 3.7392 & 8630 & 104.5620 & 0.2198 & 0.2006 & \textbf{7188} & 48.2594 & 2.3096 & \textbf{7188} & 46.6700 & 0.0000 & 0.0000 \\ 
\textbf{20-0.8-30-5} & 8175.16 & 108.0789 & 127.8169 & 1.4551 & 7942 & 108.0789 & 0.1086 & 0.0770 & \textbf{7374} & 20.4534 & 0.6175 & \textbf{7374} & 19.9750 & 0.0000 & 0.0000 \\ 
\textbf{30-0.8-15-1} & 3091.61 & 171.4778 & 66.3609 & 0.7593 & \textbf{3061} & 169.7800 & 0.0100 & 0.0000 & \textbf{3061} & 4.5098 & 0.0911 & \textbf{3061} & 4.4170 & 0.0000 & 0.0000 \\ 
\textbf{30-0.8-15-2} & 3506.00 & 160.2209 & 0.0000 & 5.1644 & 3506 & 160.2209 & 0.0139 & 0.0139 & \textbf{3458} & 11.7516 & 0.2655 & \textbf{3458} & 11.5390 & 0.0000 & 0.0000 \\ 
\textbf{30-0.8-15-3} & 5159.56 & 166.8339 & 44.5643 & 2.8985 & 5139 & 163.8840 & 0.0910 & 0.0867 & \textbf{4729} & 105.0818 & 7.0616 & \textbf{4729} & 100.5670 & 0.0000 & 0.0000 \\ 
\textbf{30-0.8-15-4} & 7312.13 & 160.7620 & 161.4134 & 3.4133 & 7283 & 159.4759 & 0.0925 & 0.0882 & \textbf{6693} & 53.8938 & 2.2803 & \textbf{6693} & 52.1000 & 0.0000 & 0.0000 \\ 
\textbf{30-0.8-15-5} & 6263.50 & 164.5370 & 113.3484 & 2.7050 & 6251 & 162.5860 & 0.0455 & 0.0434 & \textbf{5991} & 34.2898 & 1.3369 & \textbf{5991} & 33.3210 & 0.0000 & 0.0000 \\ 
\textbf{30-0.8-30-1} & 4871 & 332.1400 & 0.0000 & 9.2200 & 4871 & 330.9080 & 0.0085 & 0.0085 & \textbf{4830} & 27.9676 & 0.5595 & \textbf{4830} & 27.3360 & 0.0000 & 0.0000 \\ 
\textbf{30-0.8-30-2} & 7122.2 & 328.2900 & 182.3900 & 4.1100 & \textbf{6989} & 325.3570 & 0.0191 & 0.0000 & \textbf{6989} & 296.6414 & 21.9387 & \textbf{6989} & 279.8210 & 0.0000 & 0.0000 \\ 
\textbf{30-0.8-30-3} & 8124 & 337.1900 & 16.4300 & 33.6300 & 8112 & 321.8380 & 0.0488 & 0.0473 & \textbf{7746} & 2115.6020 & 49.0532 & \textbf{7746} & 2074.4600 & 0.0000 & 0.0000 \\ 
\textbf{30-0.8-30-4} & \textbf{8384} & 318.0600 & 0.0000 & 26.0900 & \textbf{8384} & 338.2490 & 0.0000 & 0.0000 & \textbf{8384} & 530.1420 & 15.6519 & \textbf{8384} & 520.0250 & 0.0000 & 0.0000 \\ 
\textbf{30-0.8-30-5} & 7442.8 & 321.4300 & 33.0900 & 17.8900 & \textbf{7428} & 344.3670 & 0.0020 & 0.0000 & \textbf{7428} & 162.6760 & 2.9126 & \textbf{7428} & 159.9620 & 0.0000 & 0.0000 \\ 
\textbf{30-0.8-45-1} & 6633.24 & 495.3080 & 118.1999 & 11.4544 & 6620 & 494.3174 & 0.0547 & 0.0526 & \textbf{6289} & 48.6748 & 1.2567 & \textbf{6289} & 47.7090 & 0.0000 & 0.0000 \\ 
\textbf{30-0.8-45-2} & 11150.60 & 489.6256 & 220.3763 & 15.3625 & 10975 & 489.6256 & 0.3142 & 0.2935 & \textbf{8485} & 377.5736 & 13.7328 & \textbf{8485} & 367.7370 & 0.0000 & 0.0000 \\ 
\textbf{30-0.8-45-3} & 9555.00 & 507.0021 & 399.7143 & 17.2257 & 9555 & 507.0021 & 0.2327 & 0.2327 & \textbf{7751} & 507.0248 & 20.1638 & \textbf{7751} & 495.2200 & 0.0000 & 0.0000 \\ 
\textbf{30-0.8-45-4} & 11214.00 & 492.2408 & 0.0000 & 16.3840 & 11214 & 489.3050 & 0.1906 & 0.1906 & \textbf{9419} & 2441.1600 & 31.0341 & \textbf{9419} & 2414.4100 & 0.0000 & 0.0000 \\ 
\textbf{30-0.8-45-5} & 8338.56 & 528.5251 & 155.1697 & 7.6185 & 8080 & 522.2580 & 0.0577 & 0.0249 & \textbf{7884} & 134.6612 & 1.0177 & \textbf{7884} & 133.4330 & 0.0000 & 0.0000 \\ 
 
  \hline

{\bf Avg } &     6115.6400     &    136.1477       &            &            &       6033.7111     &     135.9796 &	0.0866	& 0.0717      &     5590.1111   &    162.5785        &     &  5590.1111 & 159.2763 & 0.0000 &	0.0000
\\
\hline

\end{tabular}  
 }
\end{center}
\caption{Computational results for GRASP and VFHLB approaches for $0.8$ density instances\label{tab03}}
\end{table*}
\end{landscape}

\pagebreak

\noindent In Tables \ref{tab01}, \ref{tab02}, \ref{tab03} were used 135 instances generated by Mautonne, Labb\'{e} and Figueiredo  \cite{Mauttone2008}, whose results were published by them just for 5 instances. For these instances, the computational results suggest the efficiency of VFHLB. On average, the time spent by VFHLB was 2.31 times faster than the time spent by GRASP, being 2.954 times faster for 0.3 density networks, 3.167 times for 0.5 density networks and 0.837 times for 0.8 density networks. Also, VFHLB found all optimal solutions, while GRASP found only 44 optimal solutions. Besides that, the VFHLB also improved or equaled GRASP results for all 135 instances (91 improvements and 44 draws).\\
Another important remark is that, in Tables \ref{tab01} and \ref{tab02} VFHLB is faster than GRASP, both in the mean of Avg Times and in the mean of Best Times. Although VFHLB lose to GRASP in the mean of Avg Times and in the mean of Best Times on Table \ref{tab03}. On the other hand, GRASP finds only 26 $\%$ of the optimal solutions while, as told before, VFHLB finds all optimal solutions.\\
The experiment also showed that, at least for the instances tested, the order of the commodities set by the candidate list in the VFHLB does not change the solution obtained at the end of the algoritm, but does affect the computational time.

\subsection{Statistical Analysis}
In order to verify whether or not the differences of mean values obtained by the evaluated strategies shown in Tables \ref{tab01},\ref{tab02} and \ref{tab03} are statistically significant, we employed the Wilcoxon-Mann-Whitney test technique \cite{Hettmansperger1998}. This test could be applied to compare algorithms with some random features and identify if the difference of performance between them is due to randomness.\\
According to \cite{Hettmansperger1998}, this statistical test is used when two independent samples are compared and whenever it is necessary to have a statistical test to reject the null hypothesis, with a significance $\theta$ level (i.e., it is possible to reject the null hypothesis with the probability of ($1-\theta \times 100 \%$)). For the sake of this analysis we considered $\theta = 0.01$. The hypotheses considered in this test are:

\begin{itemize}
\item Null Hypothesis (H0): there are no significant differences between the solutions found by VFHLB and the original method;
\item Alternative Hypothesis (H1): there are significant differences (bilateral alternative) between the solutions found by VFHLB and the GRASP.
\end{itemize}

\noindent Table \ref{tab04} presents the number of better average solutions found by each strategy, for each group of instances separeted by density. The number of cases where the Null Hypothesis was rejected is also shown between parentheses.

\begin{table}[htp]
	\centering
		\begin{center}
			\begin{tabular}{ccc}
			\hline
			\multicolumn{1}{c}{\bf  Instance }   &  \multicolumn{2}{c}{{\bf Algorithms}}\\
					      \cmidrule(rl){2-3}
			\multicolumn{1}{c}{\bf   Groups}	   &   {\bf GRASP} &   {\bf VFHLB} \\
			\hline
			0.3 & 0(0) & 30(29) \\

			0.5 & 0(0) &  34(31) \\
			0.8 &	0(0)	& 43(33)\\
			\hline
			\end{tabular}  
		\end{center}
	\caption{Statistical Analysis of GRASP and DPRFLB \label{tab04}}
\end{table}

\noindent When comparing GRASP with VFHLB, we notice that almost all differences of performance ($86.91 \%$ of the tests) are statistically significant. We can also observe that the VFHLB obtained $ 100 \%$ of the best results. These results indicate the superiority of the proposed strategy.
\subsection{Complementary Analysis}

Another way to analyze the behavior of algorithms with random components is provided by time-to-target plots (TTT-plots) \cite{Aiex2006}. These plots show the cumulative probability of an algorithm reaching a prefixed target solution in the indicated running time. In TTT-plots experiment, we sorted out the execution times required for each algorithm to reach a solution at least as good as a predefined target solution. After that, the i-th sorted running time, $t_i$, is associated with a probability $p_{i} = \frac{i-0.5}{100}$ and the points $z_i = (t_i ; p_{i})$ are plotted.\\
For these experiments we tested 10 of our largest instances with a medium target (1.22 times the cost of the optimal solution). Firstly we analyze the instances with 20 nodes, followed by the analyses of instances with 30 nodes.\\

\begin{figure}[htbp]

\centering	
	    \begin{subfigure}[b]{0.48\textwidth}
                \centering
                \includegraphics[width=\textwidth]{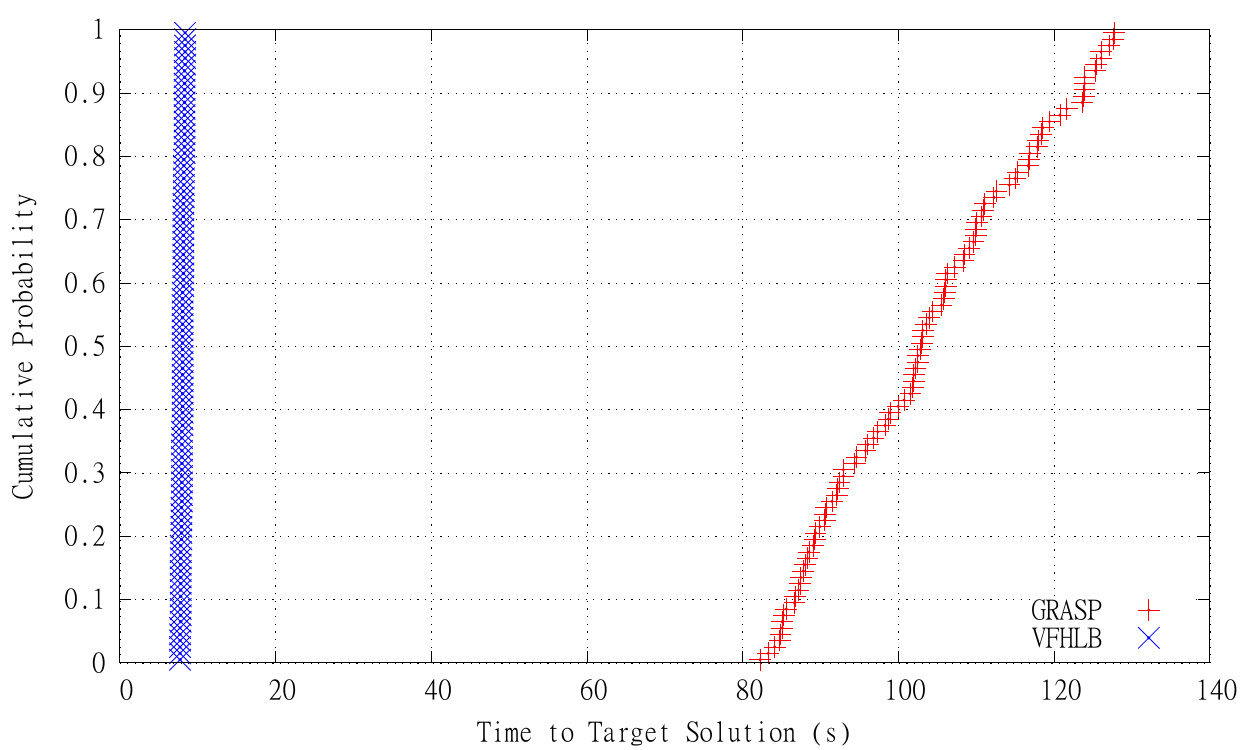}
                \caption{TTTPlot: 20-0.8-30-1}
                \label{1a}
        \end{subfigure}
    	\begin{subfigure}[b]{0.48\textwidth}
                \centering
                \includegraphics[width=\textwidth]{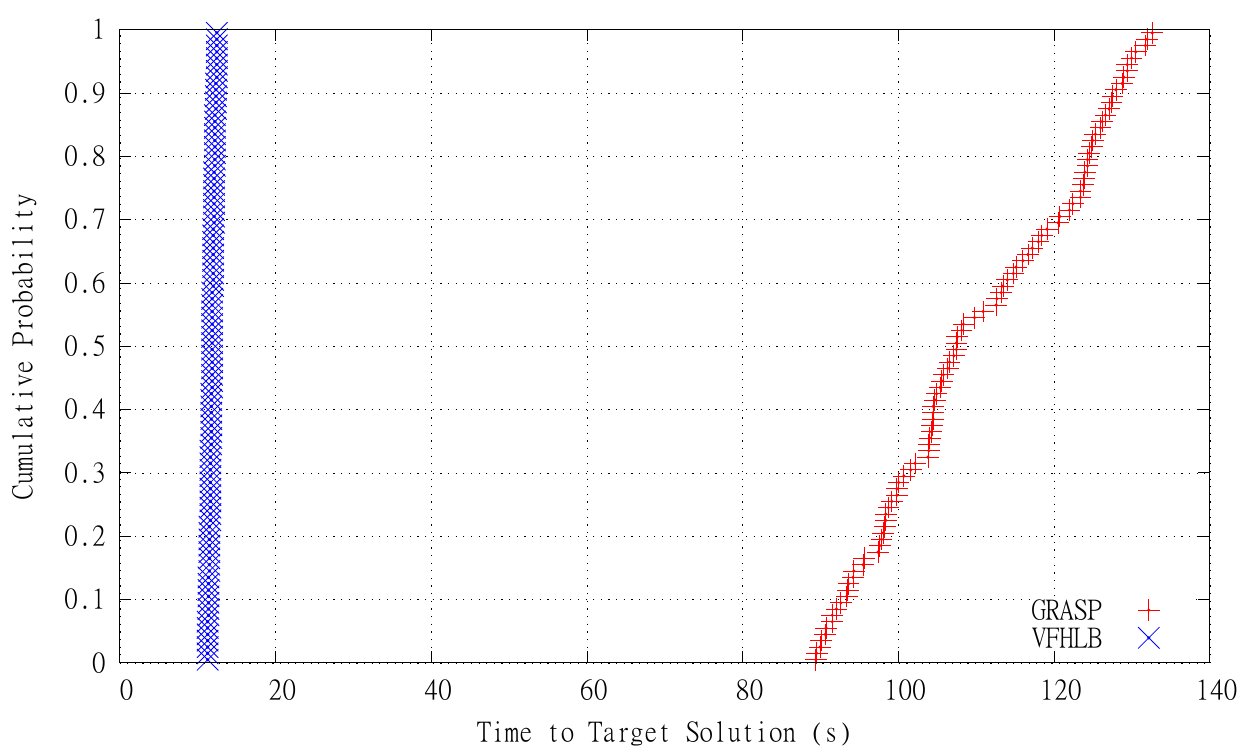}
                \caption{TTTPlot: 20-0.8-30-2}
                \label{1b}
        \end{subfigure}     
           
        \begin{subfigure}[b]{0.48\textwidth}
                \centering
                \includegraphics[width=\textwidth]{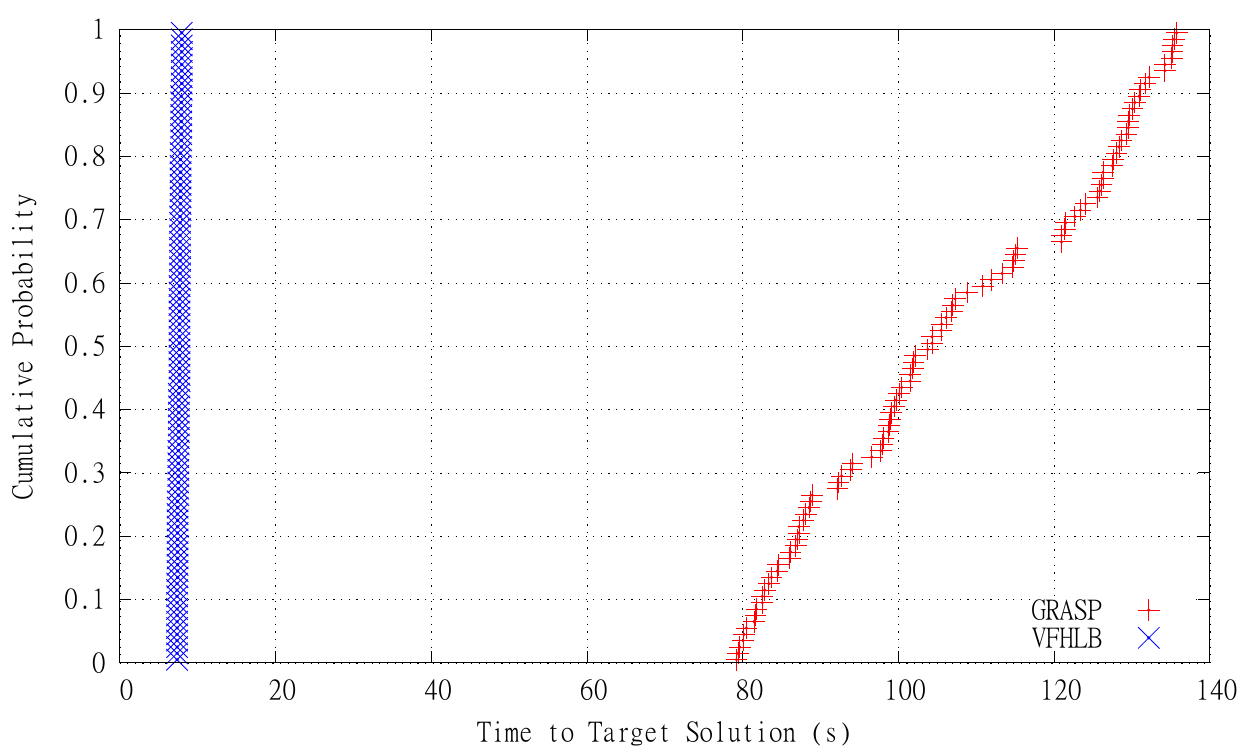}
                \caption{TTTPlot: 20-0.8-30-3}
                \label{1c}
        \end{subfigure}
        \begin{subfigure}[b]{0.48\textwidth}
                \centering
                \includegraphics[width=\textwidth]{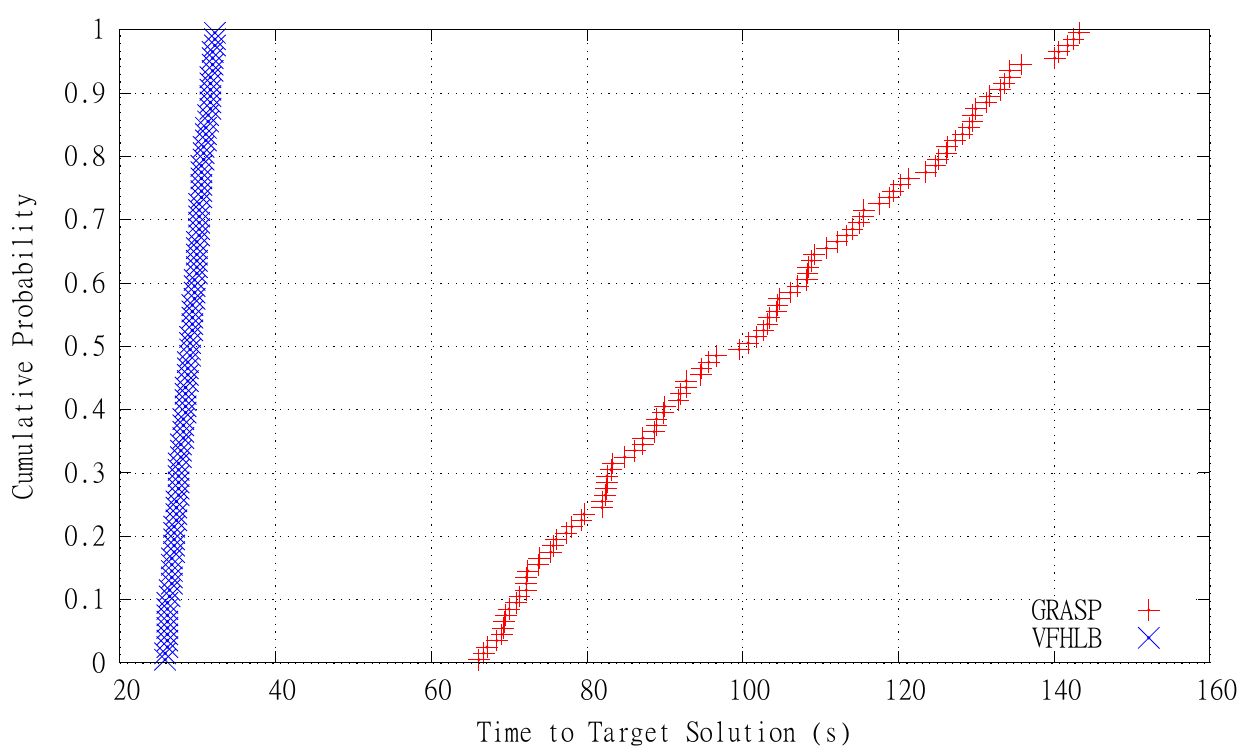}
                \caption{TTTPlot: 20-0.8-30-4}
                \label{1d}
        \end{subfigure}
        
        \begin{subfigure}[b]{0.48\textwidth}
                \centering
                \includegraphics[width=\textwidth]{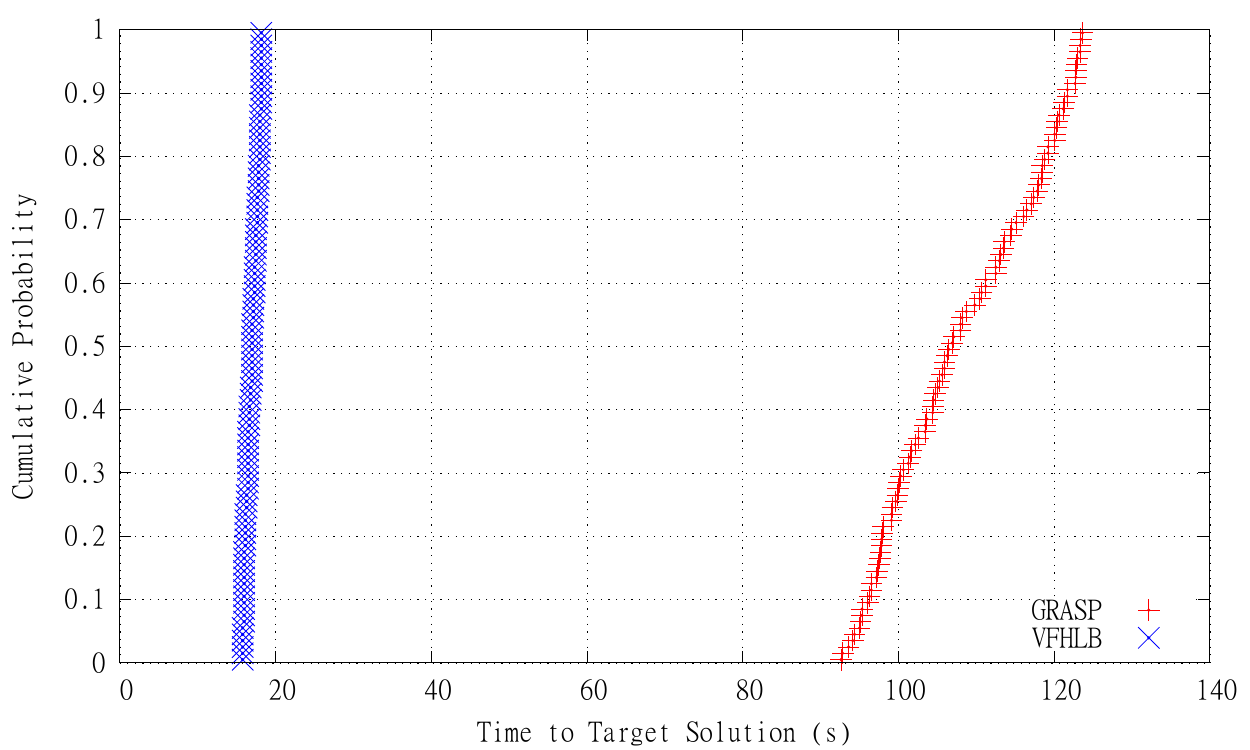}
                \caption{TTTPlot: 20-0.8-30-5}
                \label{1e}
        \end{subfigure}
		\caption{TTT Plot - 20 Nodes Instances}
\end{figure}
\pagebreak
\noindent After analyzing the behavior of the methods for the selected instances of 20 nodes, through analysis of the TTTPlot figures \ref{1a} to \ref{1e}, we conclude that the proposed strategy outperforms the GRASP, since the cumulative probability for VFHLB to find the target in less then 40 seconds is 100 $\%$, while for GRASP it is 0 $\%$.

\begin{figure}[htbp]

\centering	
	    \begin{subfigure}[b]{0.48\textwidth}
                \centering
                \includegraphics[width=\textwidth]{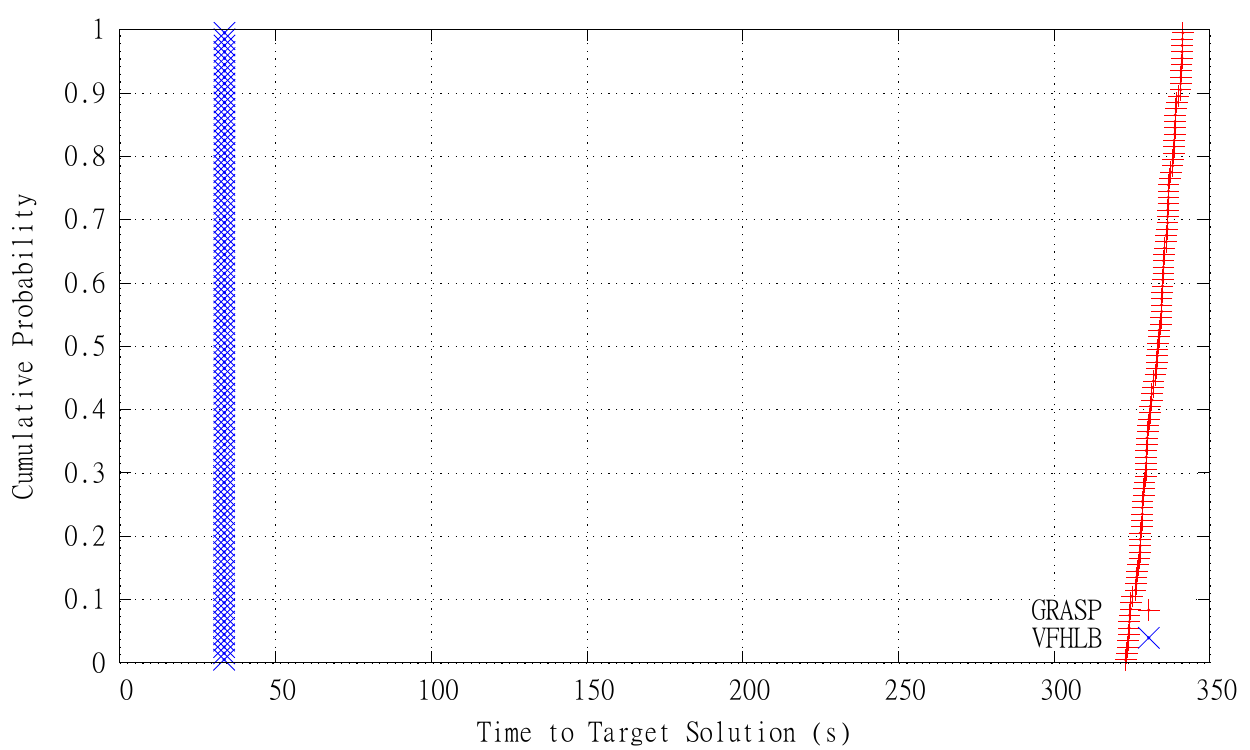}
                \caption{TTTPlot: 30-0.8-30-1}
                \label{2a}
        \end{subfigure}
    	\begin{subfigure}[b]{0.48\textwidth}
                \centering
                \includegraphics[width=\textwidth]{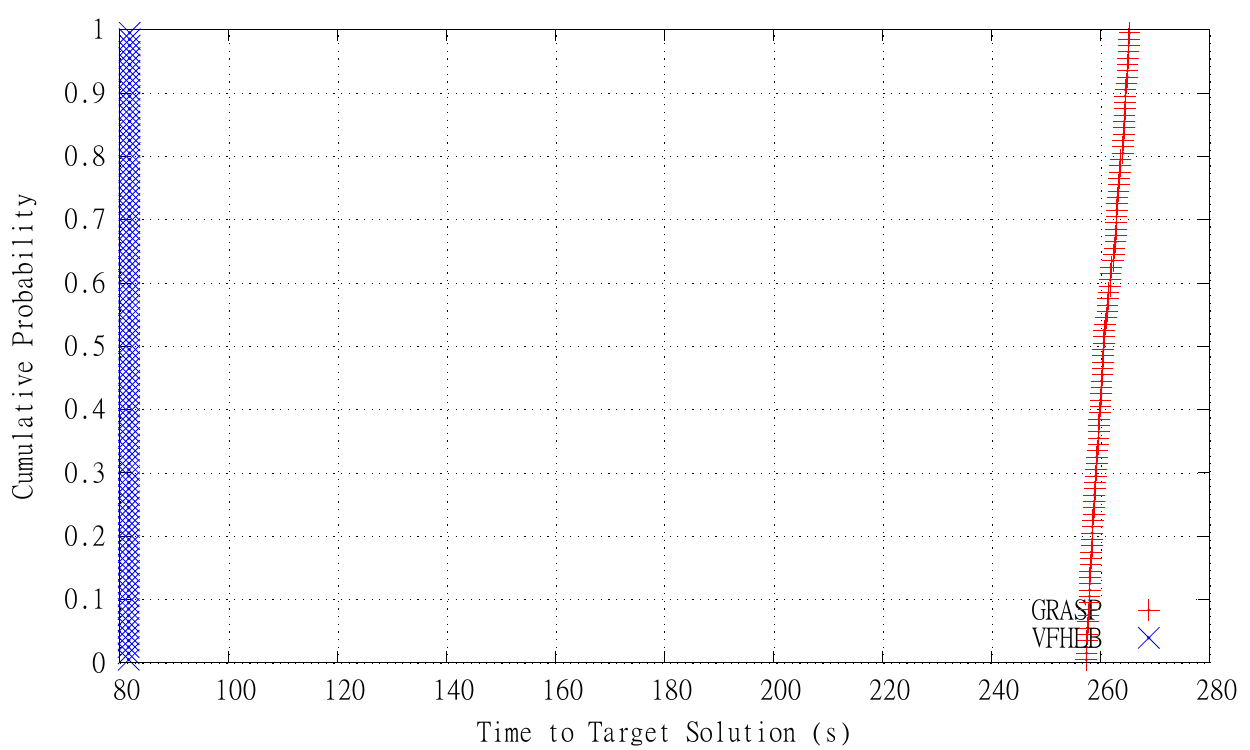}
                \caption{TTTPlot: 30-0.8-30-2}
                \label{2b}
        \end{subfigure}     
           
        \begin{subfigure}[b]{0.48\textwidth}
                \centering
                \includegraphics[width=\textwidth]{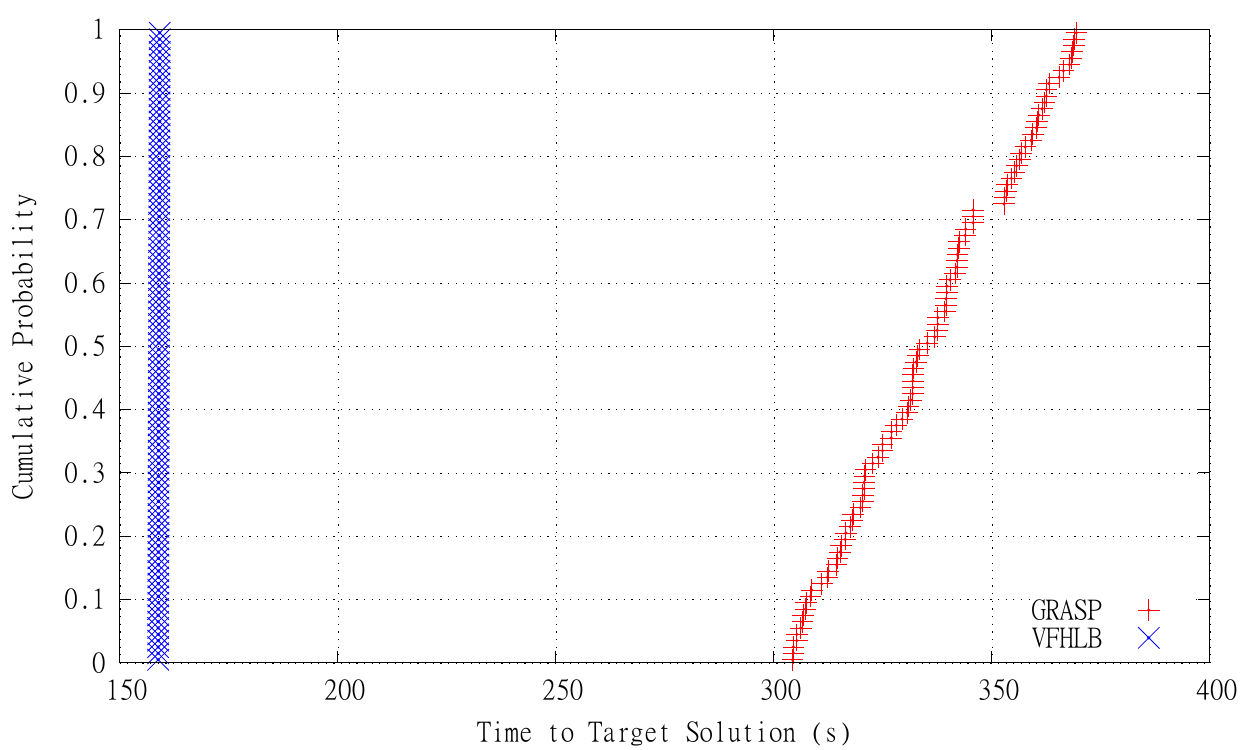}
                \caption{TTTPlot: 30-0.8-30-3}
                \label{2c}
        \end{subfigure}
        \begin{subfigure}[b]{0.48\textwidth}
                \centering
                \includegraphics[width=\textwidth]{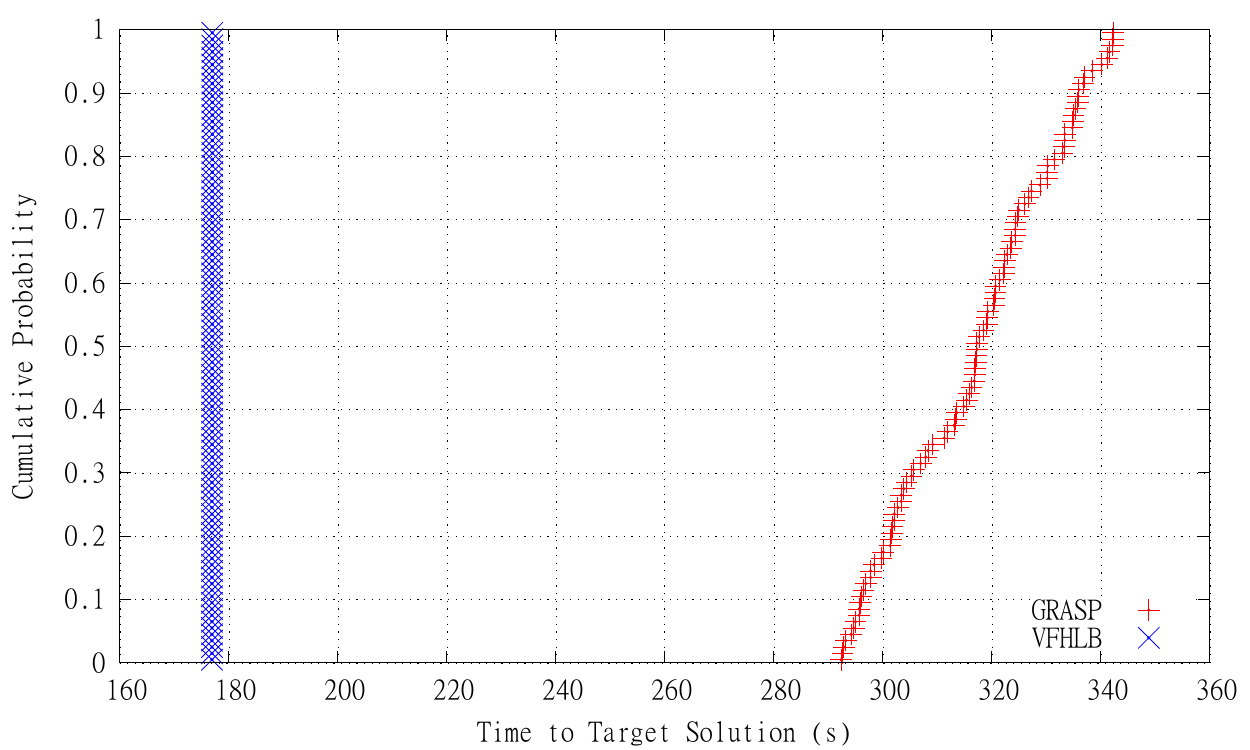}
                \caption{TTTPlot: 20-0.8-30-4}
                \label{2d}
        \end{subfigure}
        
        \begin{subfigure}[b]{0.48\textwidth}
                \centering
                \includegraphics[width=\textwidth]{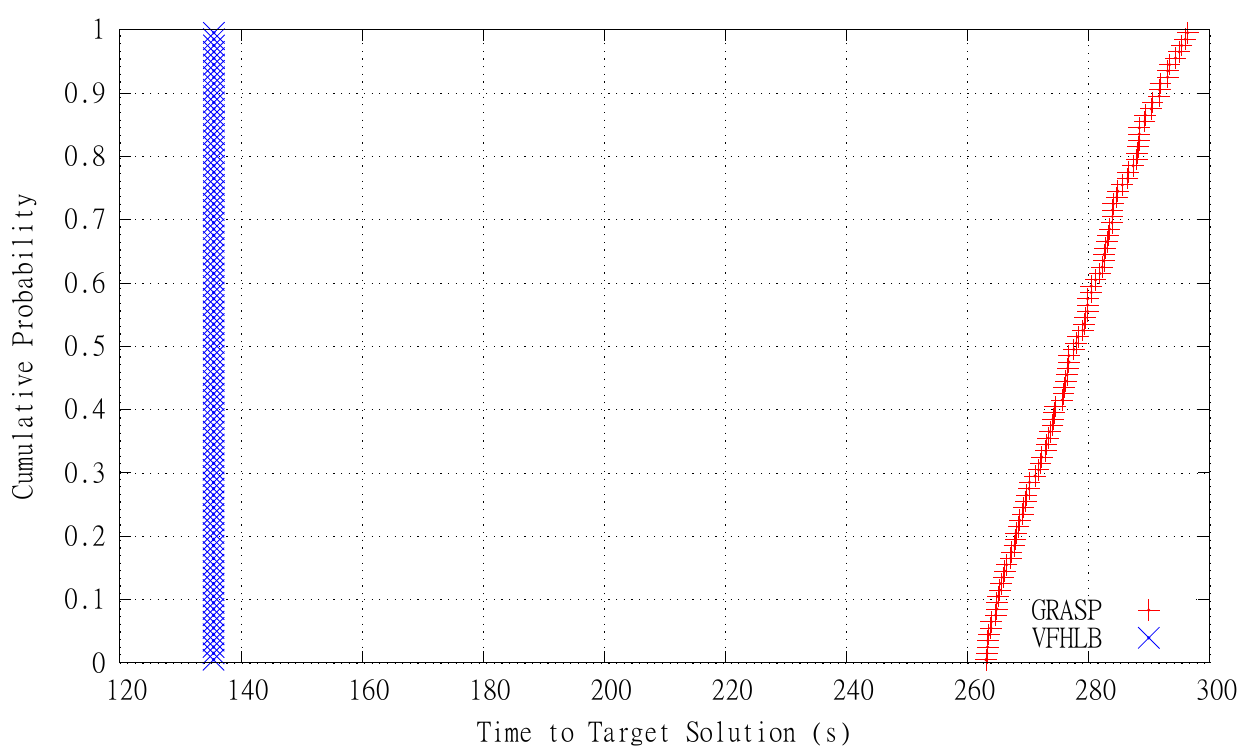}
                \caption{TTTPlot: 30-0.8-30-5}
                \label{2e}
        \end{subfigure}
		\caption{TTT Plot - 30 Nodes Instances}
\end{figure}

\noindent After analyzing the behavior of the methods for the selected instances of 30 nodes, through analysis of the TTTPlot figures \ref{2a} to \ref{2e}, we conclude that the proposed strategy outperforms the GRASP, since the cumulative probability for VFHLB to find the target in less then 180 seconds is 100 $\%$, while for GRASP it is 0 $\%$.

\section*{CONCLUSIONS}
We proposed a new algorithm for a variant of the fixed-charge uncapacitated network design problem where multiple shortest path problems were taken into consideration. In the first phase of the algorithm, the VFH is used to build a initial solution and find a lower bound. In a second moment, a Local Branching technique and a pertubation, Ejection Cycle, are applied to reduce the solution cost.\\
The proposed approach was tested on a set of instances grouped by number of nodes, graph density and number of commodities to be transported. Our results have shown the efficiency of VFHLB in comparison with the GRASP presented in \cite{Gonzalez2013}, since the proposed algorithm finds the optimal solution for all instances and presents a best average time for the majority of the instances (125 out 135).\\
As future work, we intend to work on exact approaches as Benders' Decomposition and Lagrangian Relaxation since both are very effective for similar problems, as could be seen in \cite{Bektas2007,Costa2009}.

\section*{ACKNOWLEDGEMENTS}

This work was supported by CAPES (Pedro Henrique Gonz\'alez - Process Number: BEX 9877/13-4), CNPQ (PVE Program Philippe Michelon - Process 313831/2013-0 -  Luidi Simonetti - Process 304793/2011-6) and by Laboratoire d'Informatique d'Avignon, Université d'Avignon et des Pays de Vaucluse, Avignon, France.

\bibliographystyle{elsarticle-harv}
\bibliography{Bibliografy}

\begin{thebibliography}{31}
\expandafter\ifx\csname natexlab\endcsname\relax\def\natexlab#1{#1}\fi
\expandafter\ifx\csname url\endcsname\relax
  \def\url#1{\texttt{#1}}\fi
\expandafter\ifx\csname urlprefix\endcsname\relax\def\urlprefix{URL }\fi

\bibitem[{Ahuja et~al.(1993)Ahuja, Magnanti, and Orlin}]{Ahuja:1993:NFT:137406}
Ahuja, R.~K., Magnanti, T.~L., Orlin, J.~B., 1993. {Network flows: theory,
  algorithms, and applications}. Prentice-Hall, Inc., Upper Saddle River, NJ,
  USA.

\bibitem[{{Aiex, R. M. and Resende, M. G. C. and Ribeiro, C.
  C.}(2007)}]{Aiex2006}
{Aiex, R. M. and Resende, M. G. C. and Ribeiro, C. C.}, 2007. Ttt plots: a perl
  program to create time-to-target plots. Optimization Letters 1~(4), 355--366.

\bibitem[{Amaldi et~al.(2011)Amaldi, Bruglieri, and
  Fortz}]{Amaldi2011:HTN:2040817.2040858}
Amaldi, E., Bruglieri, M., Fortz, B., 2011. {On the hazmat transport network
  design problem}. In: Proceedings of the 5th international conference on
  Network optimization. INOC'11. Springer-Verlag, Berlin, Heidelberg, pp.
  327--338.

\bibitem[{Bazaraa et~al.(2004)Bazaraa, Jarvis, and
  Sherali}]{Bazaraa:2004:LPN:1062374}
Bazaraa, M.~S., Jarvis, J.~J., Sherali, H.~D., 2004. {Linear Programming and
  Network Flows}. Wiley-Interscience.

\bibitem[{Bektas et~al.(2007)Bektas, Crainic, and Gendron}]{Bektas2007}
Bektas, T., Crainic, T.~G., Gendron, B., May 2007. Lagrangean decomposition for
  the multicommodity capacitated network design problem. In: Optimization Days
  2007.
\newline\urlprefix\url{http://eprints.soton.ac.uk/55826/}

\bibitem[{Billheimer and Gray(1973)}]{Billheimer1973}
Billheimer, J.~W., Gray, P., feb 1973. {Network Design with Fixed and Variable
  Cost Elements}. Transportation Science 7~(1), 49--74.

\bibitem[{Boesch(1976)}]{Boesch1976}
Boesch, F.~T., 1976. {Large-scale Networks: Theory and Design}, 1st Edition.
  IEEE Press selected reprint series.

\bibitem[{Boyce and Janson(1980)}]{Boyce1980}
Boyce, D., Janson, B., mar 1980. {A discrete transportation network design
  problem with combined trip distribution and assignment}. Transportation
  Research Part B: Methodological 14~(1-2), 147--154.

\bibitem[{Colson et~al.(2005)Colson, Marcotte, and Savard}]{Colson2005}
Colson, B., Marcotte, P., Savard, G., jun 2005. {Bilevel programming: A
  survey}. 4OR 3~(2), 87--107.

\bibitem[{Costa et~al.(2009)Costa, Cordeau, and Gendron}]{Costa2009}
Costa, A.~M., Cordeau, J.-F., Gendron, B., 2009. Benders, metric and cutset
  inequalities for multicommodity capacitated network design. Computational
  Optimization and Applications 42~(3), 371--392.

\bibitem[{Erkut and Gzara(2008)}]{Erkut2008}
Erkut, E., Gzara, F., jul 2008. {Solving the hazmat transport network design
  problem}. Computers \& Operations Research 35~(7), 2234--2247.

\bibitem[{Erkut et~al.(2007)Erkut, Tjandra, and Verter}]{Erkut2007}
Erkut, E., Tjandra, S.~A., Verter, V., 2007. {Hazardous Materials
  Transportation}. In: Handbooks in Operations Research and Management Science.
  Vol.~14. Ch.~9, pp. 539--621.

\bibitem[{Fischetti and Lodi(2003)}]{Fischetti2003}
Fischetti, M., Lodi, A., sep 2003. {Local branching}. Mathematical Programming
  98~(1-3), 23--47.

\bibitem[{Gonz\'{a}lez et~al.(2013)Gonz\'{a}lez, Martinhon, Simonetti, Santos,
  and Michelon}]{Gonzalez2013}
Gonz\'{a}lez, P.~H., Martinhon, C., Simonetti, L., Santos, E., Michelon, P.,
  2013. {Uma Metaheur\'{\i}stica GRASP para o Problema de Planejamento de Redes
  com Rotas \'{O}timas para o Usu\'{a}rio}. In: XLV Simp\'{o}sio Brasileiro de
  Pesquisa Operacional. Natal, pp. 1813--1824.

\bibitem[{Graves and Lamar(1983)}]{Graves1983}
Graves, S.~C., Lamar, B.~W., may 1983. {An Integer Programming Procedure for
  Assembly System Design Problems}. Operations Research 31~(3), 522--545.

\bibitem[{Hettmansperger and McKean(1998)}]{Hettmansperger1998}
Hettmansperger, T.~P., McKean, J.~W., 1998. {Robust nonparametric statistical
  methods}. CRC Press.

\bibitem[{Holmberg and Yuan(2004)}]{Holmberg2004}
Holmberg, K., Yuan, D., jan 2004. {Optimization of Internet Protocol network
  design and routing}. Networks 43~(1), 39--53.

\bibitem[{Johnson et~al.(1978)Johnson, Lenstra, and Kan}]{Johnson1978}
Johnson, D.~S., Lenstra, J.~K., Kan, A. H. G.~R., jan 1978. {The complexity of
  the network design problem}. Networks 8~(4), 279--285.

\bibitem[{Kara and Verter(2004)}]{Kara2004}
Kara, B.~Y., Verter, V., may 2004. {Designing a Road Network for Hazardous
  Materials Transportation}. Transportation Science 38~(2), 188--196.

\bibitem[{Kimemia and Gershwin(1978)}]{Kimemia1978}
Kimemia, J., Gershwin, S., 1978. {Network flow optimization in flexible
  manufacturing systems}. In: 1978 IEEE Conference on Decision and Control
  including the 17th Symposium on Adaptive Processes. IEEE, pp. 633--639.

\bibitem[{Louren\c{c}o and S.(2010)}]{Lourenco2010}
Louren\c{c}o, H., O.~M., S., T., 2010. {Handbook of Metaheuristics}. Vol. 146
  of International Series in Operations Research \& Management Science.
  Springer US, Boston, MA.

\bibitem[{{Luigi De Giovanni}(2004)}]{LuigiDeGiovanni2004}
{Luigi De Giovanni}, 2004. {The Internet Protocol Network Design Problem with
  Reliability and Routing Constraints}. Ph.D. thesis, Politecnico di Torino.

\bibitem[{Magnanti(1981)}]{Magnanti1981}
Magnanti, T.~L., jan 1981. {Combinatorial optimization and vehicle fleet
  planning: Perspectives and prospects}. Networks 11~(2), 179--213.

\bibitem[{Magnanti and Wong(1984)}]{Magnanti1984}
Magnanti, T.~L., Wong, R.~T., feb 1984. {Network Design and Transportation
  Planning: Models and Algorithms}. Transportation Science 18~(1), 1--55.

\bibitem[{Mandl(1981)}]{Mandl1981}
Mandl, C.~E., aug 1981. {A survey of mathematical optimization models and
  algorithms for designing and extending irrigation and wastewater networks}.
  Water Resources Research 17~(4), 769--775.

\bibitem[{Mauttone et~al.(2008)Mauttone, Labb\'{e}, and
  Figueiredo}]{Mauttone2008}
Mauttone, A., Labb\'{e}, M., Figueiredo, R. M.~V., 2008. {A Tabu Search
  approach to solve a network design problem with user-optimal flows}. In: V
  ALIO/EURO Conference on Combinatorial Optimization. Buenos Aires, pp. 1--6.

\bibitem[{Paraskevopoulos et~al.(2013)Paraskevopoulos, Bektaş, Crainic, and
  Potts}]{Paraskevopoulos2013}
Paraskevopoulos, D.~C., Bektaş, T., Crainic, T.~G., Potts, C.~N., 2013. {A
  Cycle-Based Evolutionary Algorithm for the Fixed-Charge Capacitated
  Multi-Commodity Network Design Problem}. Tech. rep., Centre
  interuniversitaire de recherche sur les r\'{e}seaux d'entreprise, la
  logistique et le transport, Montreal.

\bibitem[{Simpson(1969)}]{Simpson1969}
Simpson, R.~W., 1969. {Scheduling and routing models for airline systems}.
  Massachusetts Institute of Technology, Flight Transportation Laboratory.

\bibitem[{Wolsey(1998)}]{Wolsey1998}
Wolsey, L.~A., 1998. {Integer programming}. Wiley-Interscience, New York, NY,
  USA.

\bibitem[{Wong(1978)}]{Wong1978}
Wong, R.~T., 1978. {Accelerating Benders decomposition for network design}.
  Ph.D. thesis, Massachusetts Institute of Technology.

\bibitem[{Wong(1980)}]{Wong1980}
Wong, R.~T., mar 1980. {Worst-Case Analysis of Network Design Problem
  Heuristics}. SIAM Journal on Algebraic Discrete Methods 1~(1), 51--63.

\end{thebibliography}










\end{document}